\documentclass[twoside,twocolumn,9pt]{article}
\usepackage{extsizes}
\usepackage[super,sort&compress,comma]{natbib} 
\usepackage[version=3]{mhchem}
\usepackage[left=1.5cm, right=1.5cm, top=1.785cm, bottom=2.0cm]{geometry}
\usepackage{balance}
\usepackage{mathptmx}
\usepackage{sectsty}
\usepackage{graphicx} 
\usepackage{lastpage}
\usepackage[format=plain,justification=justified,singlelinecheck=false,font={stretch=1.125,small,sf},labelfont=bf,labelsep=space]{caption}
\usepackage{float}
\usepackage{fancyhdr}
\usepackage{fnpos}
\usepackage[english]{babel}
\addto{\captionsenglish}{%
  
}
\usepackage{array}
\usepackage{droidsans}
\usepackage{charter}
\usepackage[T1]{fontenc}
\usepackage[usenames,dvipsnames]{xcolor}
\usepackage{setspace}
\usepackage[compact]{titlesec}
\usepackage{hyperref}
\usepackage{grffile}
\usepackage{dcolumn}
\usepackage{bm}
\usepackage{color}

\usepackage{soul,xcolor}   
\usepackage{ulem}   
\newcommand{\nt}[1]{\textcolor{black}{#1}}

\usepackage{epstopdf}

\definecolor{cream}{RGB}{222,217,201}

\usepackage{xr}
\makeatletter
\newcommand*{\addFileDependency}[1]{
  \typeout{(#1)}
  \@addtofilelist{#1}
  \IfFileExists{#1}{}{\typeout{No file #1.}}
}
\makeatother
\newcommand*{\myexternaldocument}[1]{%
    \externaldocument{#1}%
    \addFileDependency{#1.tex}%
    \addFileDependency{#1.aux}%
}
\myexternaldocument{supplementary}

\begin{document}

\pagestyle{fancy}
\thispagestyle{plain}
\fancypagestyle{plain}{
\renewcommand{\headrulewidth}{0pt}
}

\makeFNbottom
\makeatletter
\renewcommand\LARGE{\@setfontsize\LARGE{15pt}{17}}
\renewcommand\Large{\@setfontsize\Large{12pt}{14}}
\renewcommand\large{\@setfontsize\large{10pt}{12}}
\renewcommand\footnotesize{\@setfontsize\footnotesize{7pt}{10}}
\makeatother

\renewcommand{\thefootnote}{\fnsymbol{footnote}}
\renewcommand\footnoterule{\vspace*{1pt}%
\color{cream}\hrule width 3.5in height 0.4pt \color{black}\vspace*{5pt}} 
\setcounter{secnumdepth}{5}

\makeatletter 
\renewcommand\@biblabel[1]{#1}            
\renewcommand\@makefntext[1]%
{\noindent\makebox[0pt][r]{\@thefnmark\,}#1}
\makeatother 
\renewcommand{\figurename}{\small{Fig.}~}
\sectionfont{\sffamily\Large}
\subsectionfont{\normalsize}
\subsubsectionfont{\bf}
\setstretch{1.125} 
\setlength{\skip\footins}{0.8cm}
\setlength{\footnotesep}{0.25cm}
\setlength{\jot}{10pt}
\titlespacing*{\section}{0pt}{4pt}{4pt}
\titlespacing*{\subsection}{0pt}{15pt}{1pt}

\fancyfoot{}
\fancyfoot[LO,RE]{\vspace{-7.1pt}\includegraphics[height=9pt]{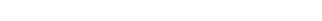}}
\fancyfoot[CO]{\vspace{-7.1pt}\hspace{13.2cm}\includegraphics{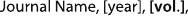}}
\fancyfoot[CE]{\vspace{-7.2pt}\hspace{-14.2cm}\includegraphics{head_foot/RF}}
\fancyfoot[RO]{\footnotesize{\sffamily{1--\pageref{LastPage} ~\textbar  \hspace{2pt}\thepage}}}
\fancyfoot[LE]{\footnotesize{\sffamily{\thepage~\textbar\hspace{3.45cm} 1--\pageref{LastPage}}}}
\fancyhead{}
\renewcommand{\headrulewidth}{0pt} 
\renewcommand{\footrulewidth}{0pt}
\setlength{\arrayrulewidth}{1pt}
\setlength{\columnsep}{6.5mm}
\setlength\bibsep{1pt}

\makeatletter 
\newlength{\figrulesep} 
\setlength{\figrulesep}{0.5\textfloatsep} 

\newcommand{\topfigrule}{\vspace*{-1pt}%
\noindent{\color{cream}\rule[-\figrulesep]{\columnwidth}{1.5pt}} }

\newcommand{\botfigrule}{\vspace*{-2pt}%
\noindent{\color{cream}\rule[\figrulesep]{\columnwidth}{1.5pt}} }

\newcommand{\dblfigrule}{\vspace*{-1pt}%
\noindent{\color{cream}\rule[-\figrulesep]{\textwidth}{1.5pt}} }

\makeatother

\twocolumn[
  \begin{@twocolumnfalse}
{\includegraphics[height=30pt]{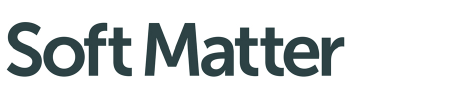}\hfill\raisebox{0pt}[0pt][0pt]{\includegraphics[height=55pt]{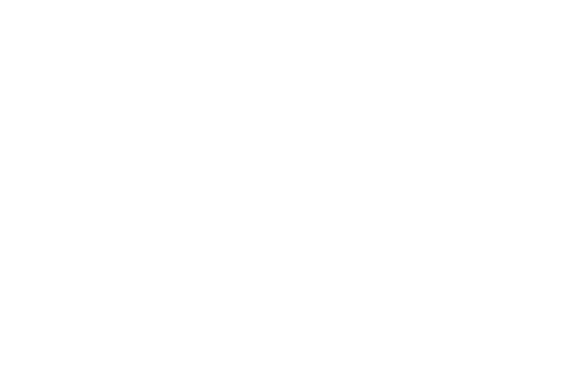}}\\[1ex]
\includegraphics[width=18.5cm]{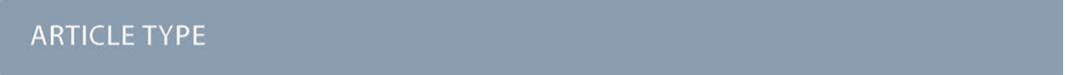}}\par
\vspace{1em}
\sffamily
\begin{tabular}{m{4.5cm} p{13.5cm} }

\includegraphics{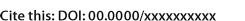} & \noindent\LARGE{\textbf{Effect of Confinement and Topology: 2-TIPS vs MIPS$^\dag$}} \\
\vspace{0.3cm} & \vspace{0.3cm} \\

 & \noindent\large{Nayana Venkatareddy,\textit{$^{a}$} Jaydeep Mandal,\textit{$^{a}$} and Prabal K. Maiti$^{\ast}$\textit{$^{a}$}} \\

\includegraphics{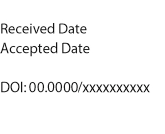} & \noindent\normalsize{2-TIPS (Two Temperature induced phase separation) refers to the phase separation phenomenon observed in mixtures of active and passive particles which are modelled using scalar activity. The active particles are connected to a thermostat at high temperature while the passive particles are connected to the thermostat at low temperature and the relative temperature difference between "hot" and "cold" particles is taken as the measure of the activity \(\chi\) of the non-equilibrium system. The study of such binary mixtures of hot and cold particles under various kinds of confinement is an important problem in many physical and biological processes. The nature and extent of phase separation are heavily influenced by the geometry of confinement, activity, and density of the non-equilibrium binary mixture. Investigating such 3D binary mixtures confined by parallel walls, we observe that, the active and passive particles phase separate, but the extent of phase separation is reduced compared to bulk phase separation at high densities and enhanced at low densities. However, when the binary mixture of active and passive particles is confined inside a spherical cavity, the phase separation is radial for small radii of the confining sphere and the extent of phase separation is higher compared to their bulk counterparts. Confinement leads to interesting properties in the passive(cold) region like enhanced layering and high compression in the direction parallel to the confining wall.  
    In 2D, both the bulk and confined systems of the binary mixture show a significant decrement in the extent of phase separation at higher densities. 
    This observation is attributed to the trapping of active particles inside the passive cluster, which increases with density. Thus the 2D systems show structures more akin to dense-dilute phase co-existence, which is observed in Motility Induced Phase Separation in 2D active systems. The binary mixture constrained on the spherical surface also shows similar phase co-existence. Our analyses reveal that the coexistent densities observed in 2-TIPS on the spherical surface agree with the findings of previous studies on MIPS in active systems on a sphere.} 

\end{tabular}

 \end{@twocolumnfalse} \vspace{0.6cm}

  ]

\renewcommand*\rmdefault{bch}\normalfont\upshape
\rmfamily
\section*{}
\vspace{-1cm}


\footnotetext{\textit{$^{a}$Centre for Condensed Matter Theory, Department of Physics, Indian Institute of Science, Bangalore 560012, India
; E-mail: maiti@iisc.ac.in}}
\

\footnotetext{\dag~Electronic Supplementary Information (ESI) available:}


\section{Introduction:}
Active matter systems are composed of self-driven particles that can convert internal free energy to some mechanical work \cite{marchetti2013hydrodynamics,toner2005hydrodynamics,romanczuk2012active,prost2015active,ramaswamy2019active,simha2002hydrodynamic,bowick2022symmetry,PhysRevLett.110.238301,D2CP04253C,D0SM00711K,Mandal_2022}. Collective  motion and high mechanical stress are some of the characteristics of such systems. These systems are ubiquitous in nature and range from a very small to large length scales, e.g. from a colony of bacteria to a flock of birds. The behaviour of such systems is studied using a variety of theoretical and computational models, the majority of which are vectorial in nature. Among many remarkable features shown by such systems, Motility Induced Phase Separation (MIPS) is one of the most important aspects.

For a non-equilibrium system with particles having different spatially varying speeds, the local density is inversely proportional to the local speed \cite{schnitzer1993theory}. Cates and Tailleur \cite{cates2015motility} theorised the density beyond which, the homogeneous distribution of a system of active particles gives rise to a spinodal decomposition into a dense and a dilute phase, the phenomena termed as Motility Induced Phase Separation (MIPS). In 2012, Fily $et. al$ 
\cite{fily2012athermal} confirmed that a system of active Brownian particles (ABPs) in the 2D suspension phase separates in a dense and dilute region from a homogeneous structure at high densities and swimming speeds. Later, Redner $et. al.$ \cite{redner2013structure} showed that, with increasing density of the system, the phase separation occurs at a lesser activity. Further experimental confirmations on MIPS were also observed \cite{buttinoni2013dynamical,ginot2015nonequilibrium}.

Besides purely active matter systems, studying a binary mixture of active and passive particles  has significant biological and industrial applications. Such systems show various features such as phase segregation \cite{mccandlish2012spontaneous,stenhammar2015activity,C8SM00222C,doi:10.1063/5.0088259} and turbulent behaviour \cite{hinz2014motility}. To model such binary mixtures without directionality, Weber $et. al.$ \cite{weber2013long} used two different diffusivities to simulate the behaviour of active and passive particles and observed phase separation between them, where 'cold' particles (particles with smaller diffusivities) form dense clusters.  Ganai $et. al $, in 2014 \cite{ganai2014chromosome} explained the chromatin separation inside the nucleus using the two-temperature model. Theoretical models \cite{grosberg2015nonequilibrium,grosberg2018dissipation,PhysRevE.101.022120} and simulations \cite{chari2019scalar,PhysRevE.107.034607,chattopadhyay2021heating,PhysRevE.107.024701,smrek2017small,smrek2018interfacial,Elismaili2022} on various soft matter systems show phase separation between active (hot) and passive (cold) particles under the two-temperature picture.

What is the effect of different confinements on such binary mixtures? What is the effect of geometry and topology?  Before trying to answer these questions in this work, we emphasize that the surface curvature is an important aspect that plays a key role in different biological phenomena such as collective cell motility in embryogenesis \cite{keller2008reconstruction} or development of corneal epithelium \cite{collinson2002clonal} or alignment direction for particle aggregation in fluid vesicles \cite{vahid2017curvature}. Surface curvature can also induce various topological defects on systems of passive nematogens \cite{fernandez2007novel,bates2008nematic,dhakal2012nematic,shin2008topological,rajendra2023packing}, which can be controlled systematically \cite{lopez2011frustrated}. An intricate interplay between the elasticity and active interaction leads to dynamics of defect structure for active nematic shells \cite{zhang2016dynamic}. In the case of colloids, structural, thermodynamical and various other properties are also dependent on the surface curvature \cite{law2018nucleation,law2020phase}. A rich array of spatiotemporal characteristics are observed in a system of active Brownian particles on spherical surface \cite{janssen2017aging}. Self-organisation of active systems on constrained geometries has also been studied \cite{ndlec1997self}.
On the other hand, the study of various condensed matter systems inside confined geometry is also of physical and biological importance \cite{lauga2006swimming,tailleur2009sedimentation,harshey2003bacterial,D2SM01562E,Binder2010,Brumby2017}. The various phenomena that follow are consequences of the interaction between the confining wall and the system, as well as the geometry of the confinement, which results in aggregation of the particles near the walls \cite{rothschild1963non,berke2008hydrodynamic,winet1984observations,kudrolli2008swarming,elgeti2009self}. The presence of wall leads to the formation of new phases which can be found only at extremely high pressure in bulk systems\cite{doi:10.1080/08927022.2013.829227}. Confined geometry affects the interaction of colloidal particles as well \cite{kreuter2013transport,vilfan2008confinement}.
Therefore, we see that there is a great influence of confinement, in general, on various kinds of soft matter systems. The geometry of the wall can also alter the ordering in various active matter properties such as MIPS\cite{C4SM00927D,Williams2022,D2SM01012G,PhysRevE.93.062605,D3SM00362K,PhysRevE.78.031409}. The active Brownian particles under confinement aggregate near the walls at low rotational diffusion rates\cite{C4SM00927D}. Binary mixtures of microscopic algae and passive colloids in microfluidic channel also show accumulation of passive particles near the boundaries\cite{Williams2022}.

In this paper, we have addressed the question of how confinement and topology can play an important role in the phase separation of the active and passive particles in a binary mixture, using the two-temperature model, the details of which are discussed in section \ref{section:simulation details}. 
We first introduce confining parallel walls in a 3D periodic system and observe that, although the active and passive particles do tend to phase separate in the confined system, the extent of phase separation is enhanced at low densities and reduced at high densities compared to the free periodic boundary/bulk cases. We also observe that for confined systems with parallel walls, the degree of phase separation reduces with increase in density. Then we confine the binary mixture in a spherical cavity, where we observe, the phase separation between active and passive particles, but the nature of phase separation is radial for small radii of the cavity. With the increase in the radius  of the spherical cavity, the bulk effect begins to set in. Interestingly, the extent of phase separation between active and passive particles in the spherical cavity is observed to be enhanced compared to the bulk periodic systems at all densities. Next, we study the effect of confinement in 2 dimensions. Interestingly, as opposed to 3D bulk systems, the 2D periodic/bulk systems show a decrement in phase separation as density is increased. 

Similar effects are observed for the 2D confined cases as well.
We also observe that for higher densities, the phase separation between active and passive particles in confined systems is reduced compared to 2D periodic systems. These phenomena are the results of increased trapping of hot particles in cold cluster.
Finally, we constrain the system on the surface of the sphere and observe that phase separation between active and passive particles is again very small 
for high packing fractions and the structure observed on the spherical surface is compared with the motility induced phase separation observed in active systems in 2 dimensions. The details of the above-mentioned results are given in  section \ref{section:results}. Finally, we draw our conclusion and shed some light on some of the future directions of the work in section \ref{section:conclusion}
\begin{figure*}[t]
    \centering
    \includegraphics[width=0.8\textwidth]{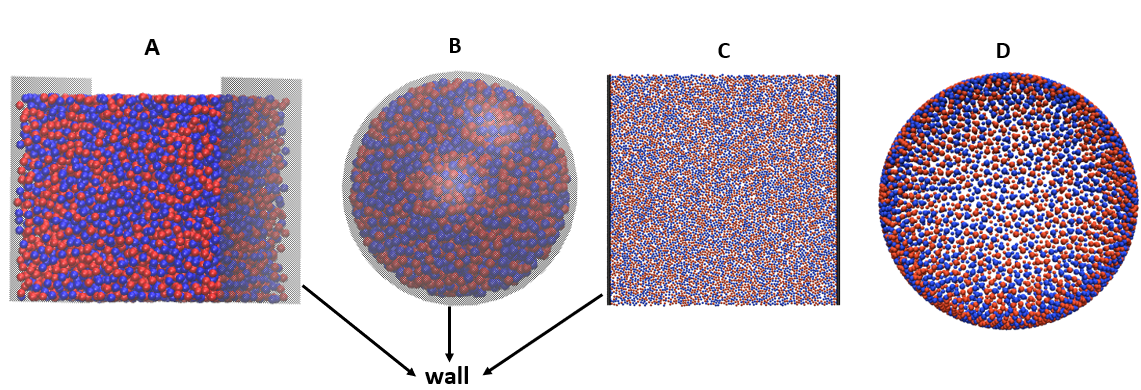}
    \caption{Initial configurations of a binary mixture of hot and cold LJ particle under various geometric confinements. A) Initial configuration of the 3D binary mixture with parallel walls perpendicular to $\hat{x}$ axis at  equilibrium when $\rho^*= 0.8$ and  $T_h^*=T_c^*=2$.  (B)Initial configuration of the 3D binary mixture under spherical confinement at  equilibrium when $\rho^*=0.8$ and  $T_h^*=T_c^*=2$ and the radius of the spherical wall is \(10\sigma\). C)Initial configuration of the 2D binary mixture with walls perpendicular to $\hat{x}$ axis at  equilibrium when $\rho^*= 0.8$ and  $T_h^*=T_c^*=2$. D)  Initial configuration of the binary mixture constrained on the surface of sphere with radius \(20\sigma\) at equilibrium when $T_h^*=T_c^*=1.0$ and $\eta=N/(16R^2) = 0.6$.} 
    \label{fig:init}
\end{figure*}

\begin{figure*}[t]
    \centering
    \includegraphics[width=0.7\textwidth]{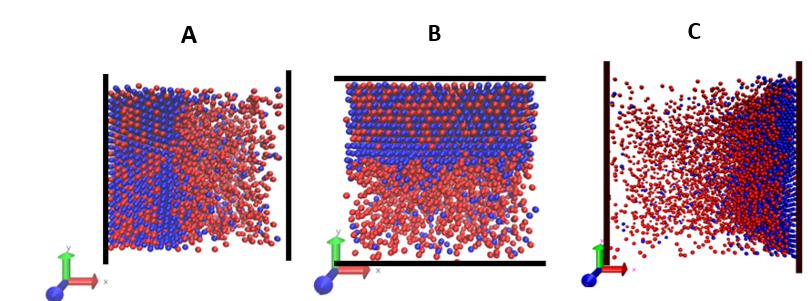}
    \caption{(A) and (B) Instantaneous configurations of the phase-separated system of hot (red) and cold (blue) particles when $\rho^*=0.8$ and  $T_h^*=80, T_c^* = 2$ in the presence of walls which are perpendicular to \(\hat{x}\) and \(\hat{y}\) directions respectively. \nt{C)} Instantaneous configuration of the phase-separated  system with walls perpendicular to $\hat{x}$ when $\rho^*=0.2$ and  $T_h^*=80, T_c^* = 2$.} 
    \label{fig:rho_0.2_config}
\end{figure*}
\section{Simulation details:}\label{section:simulation details}

We start with a system of an equal mixture of $N=8000$  hot and cold particles in a cubic periodic box as has been used in our earlier works \cite{chari2019scalar}. The particles interact via the Lennard-Jones(LJ) potential 
\begin{align} \label{wca}
    U = \begin{cases}
            4\varepsilon\left[ \left( \frac{\sigma}{r} \right)^{12} - \left( \frac{\sigma}{r} \right)^6 \right] , &r < 3.0 \sigma \\
            0 ,&r \geq 3.0 \sigma ,
        \end{cases}
\end{align}
 where $\epsilon$ carries the unit of energy, $\sigma$ is the diameter of the particles and $r$ is the distance between the particles.  We have used the reduced units throughout this work, where the various thermodynamic quantities are defined as follows:
 temperature $T^* = k_BT/\varepsilon$, pressure $P^* = P\sigma^3/\varepsilon$. 
 In our calculations, we take mass $m^*=1$, measure time in units of $(\varepsilon/m\sigma^2)^{1/2}$ and distances in units of \(\sigma\).\\
 We simulate the binary mixture of hot and cold particles under four different geometrical confinements as shown in Fig:\ref{fig:init} :\\ A) The 3D binary mixture of hot and cold LJ particles is confined by placing two parallel walls at the boundaries of simulation volume(Fig:\ref{fig:init}A), perpendicular to any one of the Cartesian directions  (say  along $\hat{x}$ axis). The interactions between the wall and particles inside the simulation volume is repulsive in nature and described by the following potential. 
 \begin{align} \label{wca-wall}
    U_{wall} = \begin{cases}
            \varepsilon\left[ \frac{2}{15}\left( \frac{\sigma}{r_w} \right)^{9} - \left( \frac{\sigma}{r_w} \right)^3 \right] +\varepsilon, & \nt{r_w < 0.4^{1/6} \sigma} \\
            0 ,& \nt{r_w \geq 0.4^{1/6} \sigma} ,
        \end{cases}
\end{align}
 where $r_w$ is the distance between the wall and a particle.\\ B)  The 3D binary mixture of hot and cold particles is confined inside spherical cavities of different radii(Fig:\ref{fig:init}B). The LJ particles interact with the wall by Weeks-Chandler-Andersen (WCA) potential.
\begin{align} \label{wca}
    U = \begin{cases}
            4\varepsilon\left[ \left( \frac{\sigma}{r_w} \right)^{12} - \left( \frac{\sigma}{r_w} \right)^6 \right] +\varepsilon, &r_w < 2^{1/6} \sigma \\
            0 ,&r_w \geq 2^{1/6} \sigma ,
        \end{cases}
\end{align}
where \(r_w\) is the distance between the particles and the wall. \\ 
C) We perform simulations of 2D binary mixtures of hot and cold particles in the presence of parallel walls at boundaries(Fig:\ref{fig:init}C). The interaction of the particles with the wall is given by eqn \ref{wca-wall}. \\ D) Finally, to study the effect of curvature, we constrain the mixture of hot and cold particles on the surface of a sphere (Fig:\ref{fig:init}D). At each timestep, all the particles obey two constraint equations:
\begin{align} \label{sphere-pos-eq}
    |\vec{r_i}| = R, \\
    \vec{r_i} \cdot \vec{v_i} = 0 \label{sphere-vel-eq}
\end{align}
 where $\vec{r_i},\vec{v_i}$ indicate the position and velocity of the $i$-th particle on the surface of the sphere of radius $R$. The origin of the coordinate system is taken at the center of the sphere.
\\

 The simulations were carried out via the LAMMPS software \cite{LAMMPS} in NVT ensemble. We have used the Nos\'e-Hoover \cite{evans1985nose} thermostat for both 3D and 2D simulations, whereas the Berendsen thermostat \cite{berendsen1984molecular} was used to control the temperature for the system on the spherical surface. The RATTLE algorithm \cite{paquay2016method} was used to constrain the particles on the surface of the sphere.  The timestep of integration was chosen to be $\delta t=0.0005$ and the time constant for the thermostats was taken as $\tau_T=100*\delta t$.\\
 Initially, all the particles are assigned the same temperature and we allow the system to equilibrate.
 The choice of the initial temperature is made such that the system remains in a fluid state in equilibrium for all the simulated densities both for 2D and 3D \cite{doi:10.1080/00268979300100411,BARKER1981226}. Accordingly, the initial temperature of the system is set at T*=2 for 3D and 2D and at T*=1 for spherical confinement (for reasons discussed in the corresponding subsection). After the equilibration, we introduce two-temperature model: connect half of the particles to a cold thermostat maintained at \(T_c^*=2\) (\(T_c^*=1.0\) for confinement on spherical surface) and the rest of the particles to a hot thermostat initially maintained at temperature  \(T_h^*=2\) (\(T_h^*=1.0\) for confinement on spherical surface). Hence, $N_{hot}, N_{cold} = N/2$ = the total number of hot and cold particles in the system. Then we increase the temperature of hot particles \(T_h^*\) from 2 to 5 to 10 .....to 80 (1 to 5 to 10....to 100 in spherical confinement). At each of the increased temperatures of hot particles \(T_h^*\), we allow the system to reach a non-equilibrium steady state  for 1 million(M) time steps. Finally, we perform a production run of another 1M time steps to obtain data for further analysis.  \\
The activity of the non-equilibrium system is measured using the relative temperature difference between the hot and cold particles, \(\chi_{imp} = \frac{T_h^*-T_c^*}{T_c^*}\). However, due to heat exchange between hot and cold particles from collisions, the effective temperature of the cold(hot) particles obtained from equipartition theorem \(T_c^{eff*}\)(\(T_h^{eff*}\)) is higher(lower) than the temperature imposed on cold(hot) thermostat \(T_c^*\)(\(T_h^*\)). Therefore the effective activity is defined as the  relative effective temperature difference between the two types of particles.
\begin{equation}
   \chi = \chi_{eff} = \frac{T_h^{eff*} - T_c^{eff*}}{T_c^{eff*}} 
\end{equation} 
We observe phase separation between hot and cold particles at high activities in the presence of any kind of  confining walls in both 3D and 2D systems and on the surface of the sphere.
 To quantify the extent of phase separation in the presence of parallel walls, we calculated the order parameter for the phase separation in the following way. We divide the simulation box into sub-cells and calculate the order parameter as \cite{chari2019scalar,chattopadhyay2021heating}
\begin{equation}\label{OP_def}
 \phi(T_{h}^*)=\frac{1}{N_{cell}}\bigg<\sum_{i=1}^{N\textsubscript{cell}} \frac{|(n^h(i)-n^c(i))|}{(n^h(i)+n^c(i))}\bigg>
\end{equation} 
where $n_h,n_c$ are the number of hot and cold particles in each of the sub-cells, $N_{cell}$ is the total number of subcells in the simulation volume and $<...>$ means the average over all steady state configurations. For the cases of particles on a spherical surface, to calculate order parameter \(\phi\), the surface is divided such that each sub-section takes almost equal areas. The polar angle $\theta \in [0,\pi]$ was divided into larger intervals near the poles of the sphere and in smaller intervals near the equator, whereas the azimuthal angle ($\phi \in [0,2\pi]$) was divided into equal parts throughout.

However, for hot and cold particles confined inside spherical walls in 3D, dividing the simulation volume into sub-cells with equal volume is challenging. So we perform cluster analysis\cite{chari2019scalar,PhysRevE.107.034607} on phase separated cold particles. We define  two particles belong to the same cluster if the distance between them is less than cut-off distance $r_c$. The cut-off distance $r_c$ is obtained from the first peak of the radial distribution function(RDF) of the cold particles. Using the above criteria of a cluster, we have calculated the normalized number of clusters of cold particles \(K_{cl}^{cold}\) and the fraction of cold particles in the largest cold cluster \(f_{cl}^{cold}\).

\section{Results:}\label{section:results}
\subsection{Confinement of 3D binary mixture by parallel walls}

\begin{figure}[!h]
    \centering
    \includegraphics[width=1\linewidth]{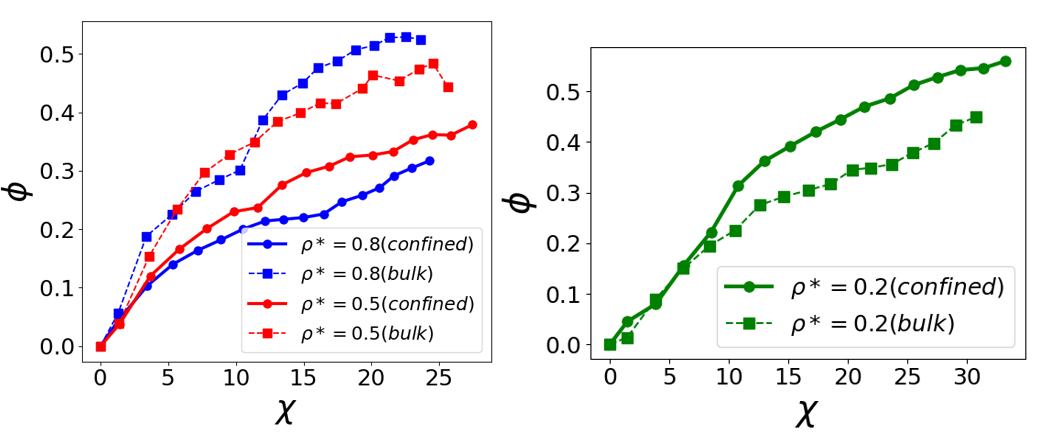}
    \caption{Plot of order parameter \(\phi\) versus activity \(\chi\) at various densities \(\rho^*\). The  solid lines are the results of the binary mixture under confinement by parallel walls while the dotted lines are the results from bulk simulations. For high densities \(\rho^*\)=0.8 and 0.5, the order parameter \(\phi\) has a lower value under confinement compared to bulk phase separation. At low density \(\rho^*\)=0.2, the order parameter \(\phi\) has higher value under confinement compared to bulk phase separation. }
    \label{fig:combined}
\end{figure}

\begin{figure}[]
    \centering
    \includegraphics[width=1.0\linewidth]{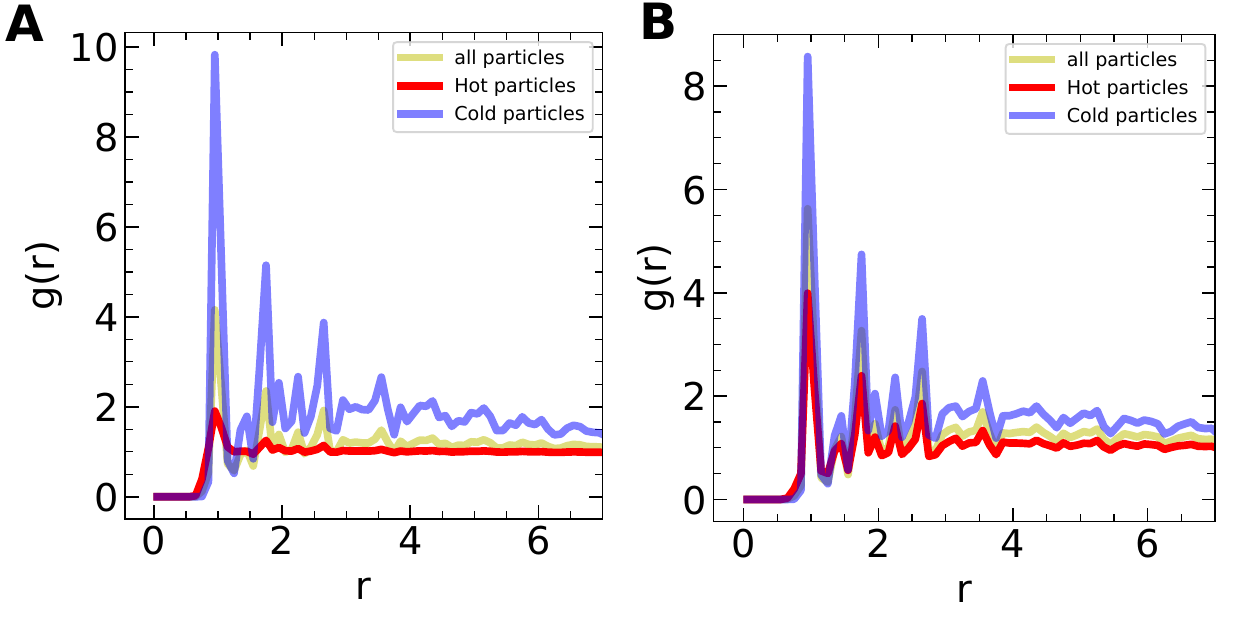}
    \caption{A) The radial distribution function of the particles for a 3D periodic system at $\rho^*=0.8,T_h^*=80.0$. The peaks in the cold cluster indicate crystalline ordering whereas, the hot particles are forming a gaseous phase. B) The radial distribution functions for the case of binary system confined by walls perpendicular to $\hat{x}$ direction, at  $\rho^*=0.8, T_h^* =80.0$. In this case, we see some peaks in the plot of hot particles as well, which indicates that a significant amount of the hot particles are trapped inside the cold cluster. }
    \label{fig:pbc-nonPBC3-gr-combined}
\end{figure}

We have observed that  in the presence of confining parallel walls, the hot and cold particles phase separate at high activities. The extent of phase separation under confinement is density dependent. The degree of phase separation is reduced compared to bulk phase separation at high densities and enhanced at low densities. \\
Fig:\ref{fig:init}A shows the initial configuration of the binary mixture with parallel walls perpendicular to \(\hat{x}\) axis when $\rho^*=N\sigma^3/V = 0.8$ ($V$ is the volume of the simulation box) and  $T_h^*=T_c^*=2$. The initial system  of hot and cold particles is well mixed. Fig:\ref{fig:rho_0.2_config}A and B show the instantaneous snapshot of phase separated system at $\rho^*=0.8$ when the temperature of hot particles is raised to $T_h^*=80$ in the presence of parallel walls perpendicular to \(\hat{x}\) and \(\hat{y}\) axis respectively. Fig:\ref{fig:rho_0.2_config}C shows the snapshot of phase separated system at $\rho^*=0.2$ when the temperature of hot particles is raised to $T_h^*=80$ and confined by walls parallel to \(\hat{x}\) axis.   \\
\textit{One of the distinguishing features of phase separation under parallel confinement is that the interface between phase-separated hot and cold particles is always parallel to the confining walls(See Fig \ref{fig:rho_0.2_config}A and B) and the phase-separated cold particles are always present in the vicinity of walls}.
To quantify the phase separation of the binary mixture under parallel wall confinement, we have calculated the order parameter \(\phi\) defined by equation \ref{OP_def} for  three simulated densities \(\rho^*\)=0.8, 0.5 and 0.2 and compared those with the results from bulk unconfined simulations. Fig \ref{fig:combined} shows the plot  of order parameter \(\phi\) versus activity \(\chi\) at various densities \(\rho^*\). The  solid lines are the results of binary mixture under confinement by parallel walls while the dotted lines are the results of bulk simulation. For all densities, we see that the order parameter \(\phi\) increases with activity \(\chi\) indicating that the hot and cold particles phase separate at high activity. We observe that for high density of \(\rho^*\)=0.8 and 0.5, the magnitude of order parameter \(\phi\) is lower under confinement compared to bulk implying a reduction in the extent of phase separation. However, at a low density of \(\rho^*\)=0.2, the value of \(\phi\) under confinement is much greater than the value of \(\phi\) from bulk simulations indicating a significant enhancement in extent of phase separation.\\
 We can  qualitatively understand the reason for the density-dependent reduction or enhancement of phase separation under parallel confinement by analyzing the composition of the phase-separated cold dense region. From Fig:\ref{fig:rho_0.2_config}A, we can see that in the phase-separated cold region, a significant number of hot particles are trapped (greater than the number of hot particles trapped in the bulk simulations without walls) at a high density of \(\rho^*\)=0.8. Under periodic boundary conditions, the cold region has two interfaces between hot and cold zones that facilitate the diffusion of hot particles trapped in the cold region. However, in the case of parallel confinement, the cold region is formed near the wall leading to only one interface between cold and hot particles. The wall prevents the escape or diffusion of the trapped hot particles resulting in significant number of hot particles in the cold region which leads to the reduction of order parameter \(\phi\) compared to bulk. This can be confirmed from the radial distribution function (RDF) of hot and cold particles for \(T_h^*\)=80 as shown in Fig:\ref{fig:pbc-nonPBC3-gr-combined}. From the RDF of bulk simulations(Fig:\ref{fig:pbc-nonPBC3-gr-combined}A) at \(\rho^*\)=0.8 and \(T_h^*\)=80, we can see that the cold particles are in the crystalline state but hot particles are in the gaseous state indicating that a negligible number of hot particles are trapped in the cold region. However, under confinement(Fig:\ref{fig:pbc-nonPBC3-gr-combined}B) at \(\rho^*\)=0.8 and \(T_h^*\)=80, the hot particles also show crystalline structure because a significant fraction of hot particles are trapped in the cold zone. The number of hot particles trapped in the phase-separated cold region decreases with density. Hence, at a low density of \(\rho^*\)=0.2, we observe an enhancement in phase separation compared with bulk. \textcolor{black}{We note that our order parameter values from the simulations (Fig:\ref{fig:combined}) are dependent on the initial configuration of the binary mixture. To obtain the results which are independent of the initial configuration, the heating rate must be very slow or the simulations must be run for very long time. The presence of wall in the system prevents the escape of trapped hot particles and hence would require extremely long simulation run time to reach its true steady state.}

\subsection{Effect of  spherical confinement:}

\begin{figure*}
    \centering
    \includegraphics[width=0.8\textwidth]{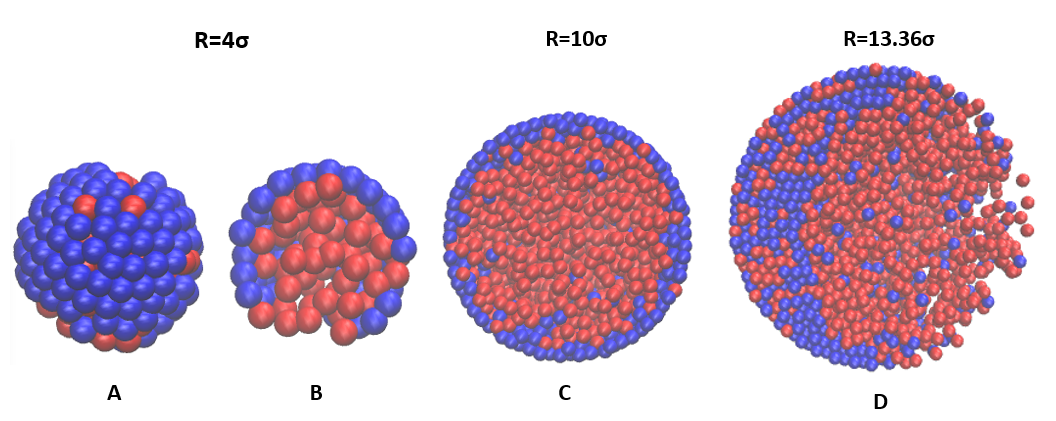}
    \caption{Instantaneous configurations of the phase separated hot and cold particles inside spherical cavities of different radii when $T_h^*=80, T_c^*=2.0$ and \(\rho^*=0.8\). A) and B) Final configuration of spherical cavity of \(R=4\sigma\) and its cross-section respectively . The phase separation is radial with cold particles near the periphery and hot particles in the interior of the sphere. C) Cross-section of final configuration of spherical cavity of \(R=10\sigma\) also shows radial phase separation similar to \(R=4\sigma\). D)Cross-section of final configuration of spherical cavity of \(R=13.36\sigma\).  For \(R=13.36\sigma\), the phase separation is not radial but takes place along a cartesian direction similar to bulk phase separation.}
    \label{fig:all_cavity_configR10R13}
\end{figure*}

In the previous section, we studied the effects of confinement of hot and cold particles by parallel walls placed at the boundary of the simulation volume. Here, we simulate the mixture of hot and cold LJ particles inside  spherical cavities of different radii to study the influence of isotropic confinement on phase separation.
We have systematically studied the effect of confining radius \(R\), density \(\rho^*\) and activity \(\chi\) on the phase separation of hot and cold LJ particles.

\subsubsection{Effect of confining radius}
We have simulated equal mixture of hot and cold LJ particles inside spherical cavities of radii \(R=4\sigma(N=214)\), \( 6\sigma(N=724)\), \( 8\sigma(N=1716)\), \(10\sigma(N=3352)\) and \(13.36\sigma(N=8000)\), while maintaining the constant density of \(\rho^*=3N/(4\pi R^3) = 0.8\). Using the same procedure mentioned in section \ref{section:simulation details}, starting from an equilibrium configuration, we maintain the temperature of cold particles at \(T_c^*=2\) and increase the temperature of hot particles from \(T_h^*=2\) to \(T_h^*=5,10,15,...80\) in steps of five. We observe that the phase separation between hot and cold LJ particles is heavily influenced by the geometry of confinement. Instantaneous snapshots of the phase-separated non-equilibrium system at different radii are shown in Fig \ref{fig:all_cavity_configR10R13}. When both hot and cold particles are at the same temperature \(T_c^*=T_h^*=2\), the particles are well mixed (Fig:\ref{fig:init}B). For \(R=4\sigma\), when the temperature of hot particles is \(T_h^*=80\), the particles undergo radial phase separation with hot particles in the interior and cold particles at the periphery of the sphere (Fig:\ref{fig:all_cavity_configR10R13}A and B). This radial phase separation is also observed for \(R=6,8,10\sigma\) at high activities (Fig:\ref{fig:all_cavity_configR10R13}C). However, for a larger radius of \(R=13.36\sigma\), when \(T_h^*=80\), the phase separation is not radial but takes place along a Cartesian direction similar to the bulk phase separation(Fig:\ref{fig:all_cavity_configR10R13}D). \textcolor{black}{Here we would like to point out that for radii \(R=4\) to \(10\sigma\) (Fig:\ref{fig:all_cavity_configR10R13}A-C), the radially phase separated cold "phase" is just 1-2 layer thick as the volume of cold region is limited by the number of particles in the finite-sized system. So it is difficult to identify the co-existing phases and their interface in the phase separated system (in contrast with phase separation under parallel walls in section 3.1 where hot and cold phases co-exist with a well defined interface). Hence we refer to the above-mentioned radial phase separation as micro-phase separation\cite{Essafri2019,doi:10.1021/jp505203t}.} As the radius of confining sphere is increased, the phase separation becomes more similar to bulk phase separation. \\

 \begin{figure*}
     \centering
     \includegraphics[width=1.0\textwidth]{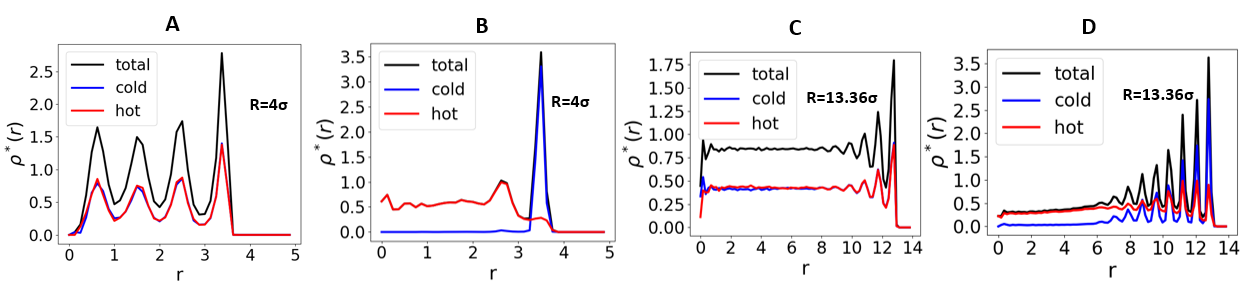}
     \caption{ Density variation of hot, cold and all particles along the radial direction. A)For \(R=4\sigma\), \(T_c^*=T_h^*=2\) and \(\rho^*=0.8\), the density of both hot and cold particles oscillate as we move along in the radial direction showing layering phenomena. B) For \(R=4\sigma\), \(T_h^*=80\) and \(\rho^*=0.8\), the density of cold particles are concentrated near the wall while the density of hot particles is uniform in the interior. C) For \(R=13.36\sigma\), \(T_c^*=T_h^*=2\) and \(\rho^*=0.8\), the density oscillations are present only near the wall and the density is constant in the interior due to the large radius of the sphere. D) For \(R=13.36\sigma\), \(T_h^*=80\) and \(\rho^*=0.8\), we see that cold particles don’t show complete radial phase separation and are present in radius range \(6< r < 13.36\), indicating phase separation takes place along the cartesian direction. }
     \label{fig:part_dist_R4_Th2_Th80-combined}
 \end{figure*}

\begin{figure*}
    \centering
    \includegraphics[width=0.8\textwidth]{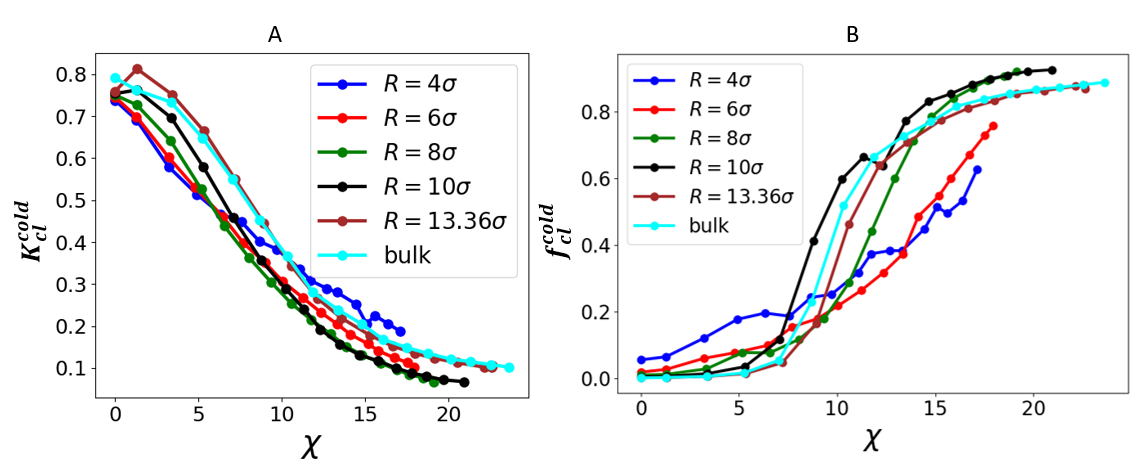}
    \caption{ A) Normalized number of clusters of cold particles \(K_{cl}^{cold}\) versus activity \(\chi \) for different values of $R$ when \(\rho^*=0.8\) . As the activity \(\chi\) increases, the number of clusters of cold particles \(K_{cl}^{cold}\) for all radii decreases. B) Fraction of cold particles in the largest cluster \(f_{cl}^{cold}\) versus activity \(\chi \) for different values of $R$ when \(\rho^*=0.8\). The size of the largest cluster increases with activity \(\chi\) for all radii. The size of the largest cluster increases as we increase the radius upto \(R=10\sigma\) and then decreases slightly for \(R=13.36\sigma\).  }
    \label{fig:ncl}
    \end{figure*}
 To understand the arrangement of hot and cold particles inside the sphere, we have calculated the density variation \(\rho^*(r)\) in the radial direction  $r$ from the center of the sphere to the wall (Fig:\ref{fig:part_dist_R4_Th2_Th80-combined}). The sphere is divided into spherical shells along the radial direction and the density of hot and cold particles is calculated in each shell. The density of both hot and cold particles oscillate along the radial direction for \(R=4\sigma\) when \(T_c^*=T_h^*=2\) and \(\rho^*=0.8\) (Fig:\ref{fig:part_dist_R4_Th2_Th80-combined}A) and indicates layering of the particles. Such layering of particles leading to density oscillation is well studied in equilibrium systems\cite{doi:10.1063/1.454434,PhysRevLett.96.177804,doi:10.1080/00268976.2013.781694}. The magnitude of density oscillations increases with density and decreases with increase in confining volume. Fig:\ref{fig:part_dist_R4_Th2_Th80-combined}B shows density variation for \(R=4\sigma\) when \(T_h^*=80\) where we can clearly see that cold particles are present near the wall and hot particles are present in the interior. Due to the high temperature of hot particles layering is absent in the interior. Also, we can see that the density of cold particles near the periphery is very high (higher than the density of phase-separated cold region in bulk as reported in our previous work\cite{chari2019scalar}) and the density of the hot particles is lower than the average density. Fig:\ref{fig:part_dist_R4_Th2_Th80-combined}C and D show the density variation for \(R=13.36\sigma\) and \(\rho^*=0.8\) when  \(T_h^*=2\) and \(T_h^*=80\) respectively. For a large radius of \(R=13.36\sigma\) in equilibrium, the density of both hot and cold particles is constant in the interior and oscillates only near the wall. When \(T_h^*=80\), the phase separation is not radial as we can see that cold particles are present in the range \(6\sigma<r<13.36\sigma\). Also the region with phase separated cold particles shows layering due to the high density of phase separated cold particles.    \\
The phase separated cold particles in the binary mixture always seem to aggregate near the wall as if there was an effective attraction between the cold particles and the wall. This attraction between the cold particles and the wall can be quantified by using the mean radial position of cold particles(Refer Fig:S1 and Section I in Supplementary Information(SI)) which increases as the activity \(\chi\) of the non-equilibrium system is increased. For radius \(R=4\sigma\) to \(10\sigma\), the 3D phase separation of hot and cold particles leads to a 2D spherical shell of cold particles where the particles are arranged in 2D hexagonal lattice \cite{GIARRITTA1993649}. For radius \(R=13.36\sigma\), the phase-separated cold cluster has particles arranged in HCP and FCC lattice similar to the bulk phase separation. Further details on the structure of phase separated cold particles are given in Section II of the SI.  \\
The hot and cold particles inside a spherical cavity phase separate at high activities for all radii even though the nature of phase separation depends on the radius. The hot particles due to their high temperature, force the cold particles to aggregate and form clusters. To study the number and size of the cold clusters, we perform cluster analysis of cold particles as defined in section II. Fig:\ref{fig:ncl}A and B give the plot of the normalized number of clusters of cold particles \(K_{cl}^{cold}\) (normalized by the total number of cold particles \(N_{cold}\)) versus activity \(\chi\) and plot of the fraction of cold particles in the largest cluster \(f_{cl}^{cold}\), defined as  $f_{cl}^{cold} = n_{cl}^{cold}/N_{cold}$ (where \(n_{cl}^{cold}\) is the number of cold particles in the largest cluster), versus activity \(\chi \) respectively  for different values of $R$. Initially for low activity, the number of  clusters of cold particles \(K_{cl}^{cold}\) is large indicating that the cold particles are spread out in the simulation volume showing no phase separation. As the activity \(\chi\) is increased, the number of  clusters of cold particles \(K_{cl}^{cold}\) decreases with activity \(\chi\) as small clusters join to form larger clusters for all radii \(R\). We see that \(K_{cl}^{cold}\) decreases with an increase in radii from \(R=4\sigma\) to \(10\sigma\) even below the value for bulk phase separation. So, the spherical confinement of hot and cold particles enhances phase separation compared to bulk phase separation. The plot of \(K_{cl}^{cold}\)  for \(R=13.36\sigma\) closely follows the plot of bulk phase separation. The fraction of cold particles in the largest cluster \(f_{cl}^{cold}\) is proportional to the size of the largest cold cluster. The fraction of cold particles in the largest cluster \(f_{cl}^{cold}\) increases with  activity \(\chi \) for all radii which implies that the size of the cold cluster grows with activity. The size of the cold cluster also increases as we increase the radius from \(R=4\sigma\) to \(10\sigma\) even surpassing the bulk phase separation. For \(R=13.36\sigma\), the size of the cluster is reduced compared to its preceding radius(\(10\sigma\)) but closely follows the value of bulk phase separation.
\begin{figure*}
    \centering
    \includegraphics[width=1.0\textwidth]{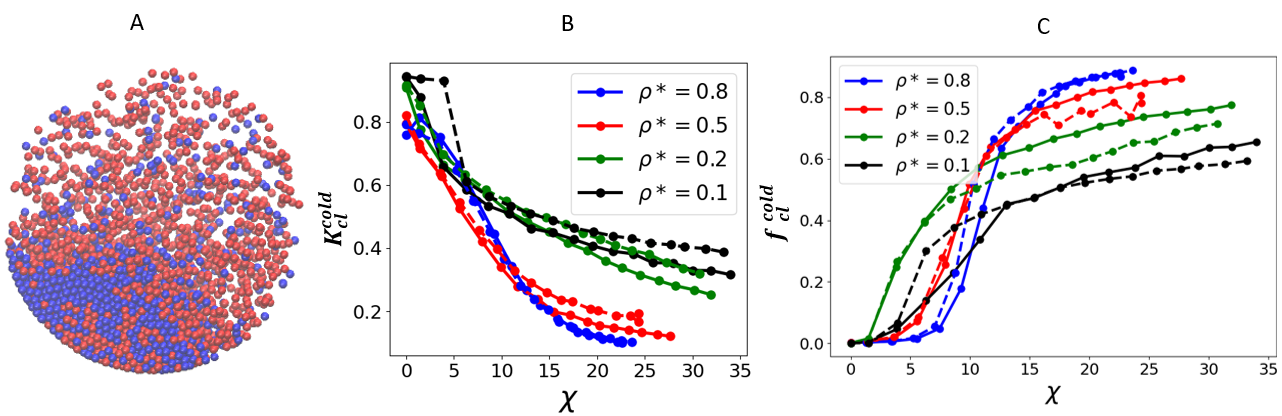}
    \caption{A)Instantaneous snapshot of the final configuration of the system of N=8000 hot and cold particles inside spherical cavity when $T_h^*=80$ at density \(\rho^*=0.2\). We can clearly see that the hot and cold particles phase separate. B)Plot of normalized number of clusters of cold particles \(K_{cl}^{cold}\) versus activity \(\chi \) for different densities $\rho^*$. The solid lines are plots for spherically confined systems and the dashed lines are plots for bulk systems. As the activity \(\chi\) increases, the number of clusters of cold particles \(K_{cl}^{cold}\) decreases for all densities. The number of clusters of cold particles \(K_{cl}^{cold}\) in spherically confined systems is always lower than the bulk systems indicating that the extent of phase separation is enhanced in spherically confined systems at all densities compared to their bulk counterparts. C)Plot of fraction of cold particles in the largest cluster \(f_{cl}^{cold}\) versus activity \(\chi \) for different densities $\rho^*$. We can clearly see that the size of the cluster is higher in spherically confined systems at all densities compared to their bulk counterparts. }
    \label{fig:den0.2}
    \end{figure*}
\subsubsection{Effect of density on phase separation inside spherical cavity.}
  We find that for all densities under spherical confinement, the phase separation is enhanced under confinement compared to bulk simulations. We simulate a fixed number of hot and cold particles (N=8000) inside spherical cavities at densities \(\rho*=0.8 (R=13.36\sigma)\), \(0.5(R=15.63\sigma)\), \(0.2(R=21.21\sigma)\) and \(0.1(R=26.72\sigma)\). Since the number of particles is fixed, the radius of the sphere is varied to obtain desired density. We observe phase separation at high activities for all densities (Fig:\ref{fig:den0.2}A). However, due to the large radii of  confining spheres, the phase separation is not radial but takes place along a cartesian direction  similar to the bulk phase separation. Again, to study the number and size of the cold clusters, we perform cluster analysis on cold particles as defined in section II. Fig:\ref{fig:den0.2}B and C show plot of normalized number of clusters of cold particles \(K_{cl}^{cold}\) versus activity \(\chi \)  and plot of fraction of cold particles in the largest cluster \(f_{cl}^{cold}\) versus activity \(\chi \) respectively  for different densities. The results of spherically confined systems (solid lines) are compared with bulk systems (dashed lines) at all densities.  As the activity \(\chi\) increases, the number of clusters of cold particles \(K_{cl}^{cold}\) for all densities decreases. The number of clusters of cold particles \(K_{cl}^{cold}\) in spherically confined systems is always lower than the bulk systems. We conclude that confining a non-equilibrium system of hot and cold particles inside a sphere enhances the extent of phase separation compared to their bulk counterparts. The  fraction of cold particles in the largest cluster \(f_{cl}^{cold}\)  is also higher in spherically confined systems compared to their bulk counterparts. Under spherical confinement, phase separation is enhanced even at high densities where we observe a reduction in the extent of phase separation under parallel walls. Due to the geometry of spherical confinement, the phase-separated hot and cold particles have a larger interface(See Fig:\ref{fig:all_cavity_configR10R13}B, C, D and \ref{fig:den0.2}A) relative to the interface under parallel confinement. So, the number of trapped hot particles in the cold region is lower in spherically confined systems compared to parallel confinement, which leads to an increase in the extent of phase separation.

\begin{figure*}
    \centering
    \includegraphics[width=0.8\linewidth]{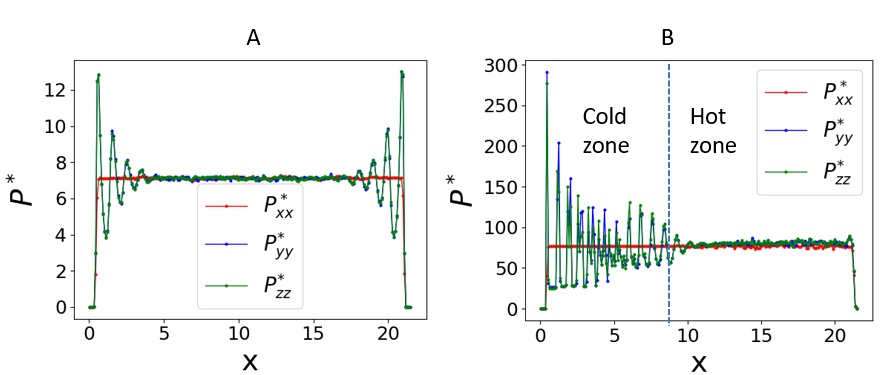}
    \caption{ A) Variation of normal pressure \(P_N\)=\(P_{xx}^*\) and tangential pressure \(P_T\)=\(P_{yy}^*\)=\(P_{zz}^*\) along \(\hat{x}\) direction perpendicular to parallel walls when \(\rho^*=0.8\) and \(T_c^*=T_h^*=2\). B)Variation of normal pressure \(P_N\)=\(P_{xx}^*\) and tangential pressure \(P_T\)=\(P_{yy}^*\)=\(P_{zz}^*\) along \(\hat{x}\) direction perpendicular to parallel walls when \(\rho^*=0.8\) and \(T_h^*=80\). We can see that in the phase-separated cold region the tangential pressures \(P_{yy}^*\) and \(P_{zz}^*\) are significantly higher than the normal pressure \(P_{xx}^*\) indicating that the cold particles are under compression.  }
    \label{fig:press}
    \end{figure*}
\subsection{Pressure anisotropy in confined systems}
The pressure variation across the interface of phase-separated hot and cold particles has been studied in bulk systems \cite{PhysRevE.107.024701,chattopadhyay2021heating,PhysRevE.107.034607,Elismaili2022} which reveal that the high kinetic pressure of the phase-separated hot particles is balanced by the high virial pressure of the dense cold clusters. Also, pressure anisotropy has been studied in bulk\cite{PhysRevE.107.024701,Elismaili2022} systems which reveal that in the cold region of the phase-separated system, the value of tangential pressure \(P_T\) is always lower than the normal pressure \(P_N\) (See Fig:4 in SI for bulk results of binary mixture of LJ particles). In case of confined systems with either parallel or spherical walls, despite the repulsive interaction of the wall with the particles, the cold particles at high activity aggregate near the walls as if there is an effective attraction between the cold particles and the wall. This attraction is also evident from the fact that the cold particles under confinement reach densities greater than their bulk counterparts. So, we investigate the effect of confinement on  the tangential and normal pressures of the binary mixture in the presence of parallel walls at the boundaries of the simulation volume. The simulation volume is divided into sub-volumes along the direction perpendicular to the wall which is along the $\hat{x}$ axis. The component of pressure tensor \(P_{\alpha \beta}\) \cite{galteland2021nanothermodynamic,galteland2022defining,doi:10.1063/5.0132487,doi:10.1080/102866202100002518a}in the \(i^{th}\) sub-volume is given by
\begin{equation}
    P_{\alpha \beta}(i)=P_{\alpha \beta}^k(i) + P_{\alpha \beta}^c(i)
\end{equation}
where \(P_{\alpha \beta}^k\) is the kinetic part and \(P_{\alpha \beta}^c\) is the configurational or virial part of the pressure tensor.
\begin{equation}
    P_{\alpha \beta}^k(i)= \frac{1}{V(i)}\sum_{j \in V(i)}mv_\alpha^j v_\beta^j
\end{equation}
\begin{equation}
    P_{\alpha \beta}^c(i)=\frac{1}{V(i)}\int_{C_{jk} \in V(i)}f_\alpha^{jk} dl_\beta
\end{equation}
where \(j \in V(i)\) implies summation over all the particles in sub-volume \(V(i)\) and \(v_\alpha^j\) is the velocity of \(j^{th}\) particle along \(\alpha\) direction. \(f_\alpha^{jk}\) is the force between \(j^{th}\) and \(k^{th}\) particles along \(\alpha\) direction, \(l\) is a point on contour \(C_{jk}\) connecting particles \(j\) and \(k\) and \(C_{jk} \in V(i)\) implies that only particles \(j\) and \(k\) contribute to the integral if a part of contour \(C_{jk}\) lies in volume \(V(i)\). 
Fig:\ref{fig:press} A) and B) shows the variation of normal pressure \(P_N\)=\(P_{xx}^*\) and tangential pressure \(P_T\)=\(P_{yy}^*\)=\(P_{zz}^*\) along \(\hat{x}\) direction perpendicular to parallel walls at \(\rho^*=0.8\) when \(T_c^*=T_h^*=2\) and \(T_h^*=80\) respectively. The condition of mechanical equilibrium \(\nabla \cdot P=0\) requires the normal component of pressure \(P_N\) to be constant. In equilibrium, when \(T_c^*=T_h^*=2\), we see that normal component \(P_{xx}^*\) is constant along \(\hat{x}\) axis but the tangential pressures \(P_{yy}^*\) and \(P_{zz}^*\) oscillate near the walls due to layering effect. When  \(T_h^*=80\), normal component \(P_{xx}^*\) is constant as dictated by the condition of mechanical equilibrium. The tangential pressures \(P_{yy}^*\) and \(P_{zz}^*\) in the cold region show oscillations of high amplitude(due to the high density of cold region layering effect is very much enhanced ) which exceeds the normal pressure \(P_{xx}^*\) by nearly order of magnitude. As mentioned before, the strong effective attraction of cold particles with the wall is also reflected in the pressure calculation, where in contrast to bulk, tangential pressure in the cold region is much greater than the normal pressure inducing strong compression in the cold regions perpendicular to the direction of the wall.

\subsection{Effect of confinement in 2 dimensional plane} \label{subsec:Effect of confinement in 2 dimensional plane}

\begin{figure}
    \centering
    \includegraphics[width=0.9\linewidth]{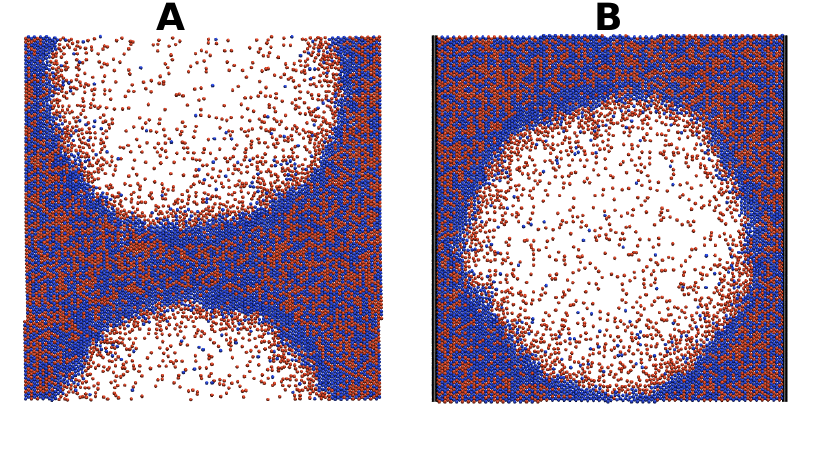}
    \caption{Snapshots of the non equilibrium steady states at \(T_h^* = 80.0, T_c^* = 2.0\) at \(\rho^* = N/(L^2) =  0.2\) for A) 2D bulk and  B) \textit{x-confined} system, where $L = $ length of the 2D plane. Formation of a single large cold cluster with hot particles trapped inside can be observed in the configurations. The red and blue particles are the hot and cold ones respectively.}
    \label{fig:2Dconfig-snapshots}
\end{figure}

We also study the effect of dimension and confinement on the phase separation of such binary mixtures. In 3D systems, we observed that for low density ($\rho^* = 0.2$), the phase separation increases with parallel confinement (wall normal to the edges along $\hat{x}$ direction), whereas the phase separation reduces with confinement for the high density ($\rho^* = 0.8$). To examine the effect of dimensions with similar parallel confinements, we have simulated a binary mixture of $N=8000$ particles on a 2D plane (square) with different densities $\rho^* = N/(L^2)$, where $L $ is the length of the plane. We have studied three cases: i) Periodic plane (2D bulk) ii) Plane with parallel walls in x direction (normal to $\hat{x}$ direction 
at the edges of the simulation region, implemented by a repulsive potential described by eqn \ref{wca-wall}) (\textit{x-confined}) and iii) Plane with parallel walls in both directions ($\hat{x}$ and $\hat{y}$) (\textit{xy-confined}). Each of these three confined systems was simulated at three different densities a) $\rho^* = 0.2,$ b) $\rho^* = 0.5$, and c) $\rho^* = 0.8$. The densities and other parameters for the simulations were chosen such that the initial configuration assumes a fluid phase.

\begin{figure}[]
    \centering
    \includegraphics[width=1.0\linewidth]{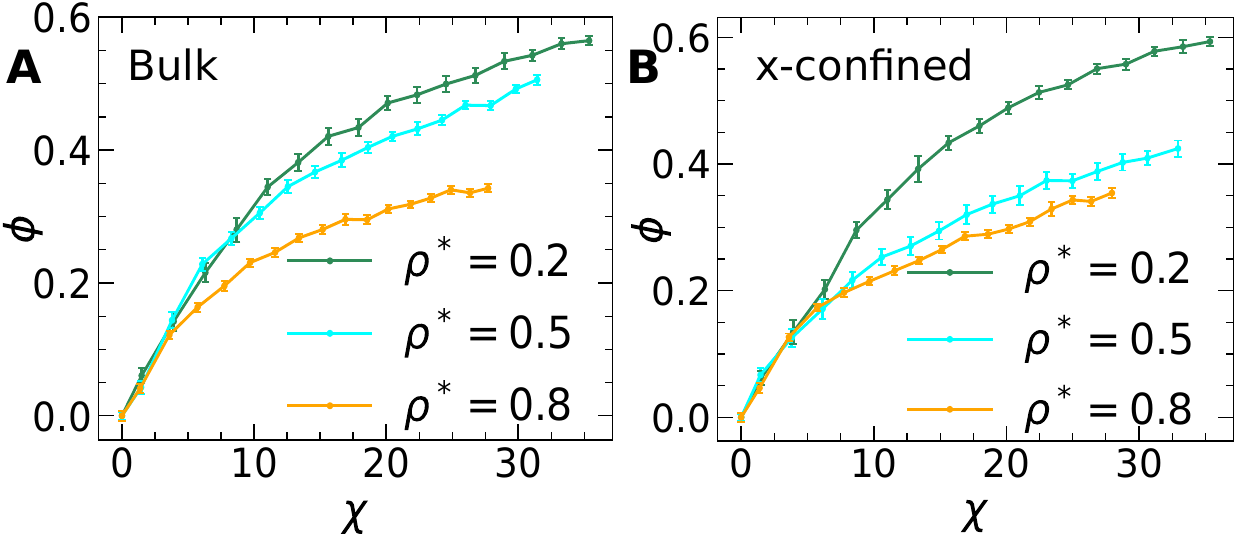}
    \caption{The effect of density on phase separation at specific confinement in 2D- Plot of order parameter \(\phi\) as a function of activity \(\chi\) at different densities, $\rho^* = 0.2, 0.5$, and $0.8$: for A) periodic plane (2D bulk), and B) \textit{x-confined} system. The extent of phase separation reduces with an increase in the density of the system for both the cases.}
    \label{fig:S_v_chi_fix-wall-diff-rho-combined}
\end{figure}

In equilibrium, at $T_c^* = 2.0$, the particles are uniformly distributed (Fig:\ref{fig:init}(C)). When two temperature scalar activity is introduced, the hot and cold particles phase separate for both the unconfined  and confined systems (see Fig:\ref{fig:2Dconfig-snapshots}A-B). We find that the phase separation phenomenon is indeed affected by the dimension of the space. In contrast to 3D bulk results, in the 2D periodic case (2D bulk), we observe that the extent of phase separation reduces with density, as is evident from the saturation values of the order parameter, as shown in Fig:\ref{fig:S_v_chi_fix-wall-diff-rho-combined}(A). Similar findings are made for \textit{x-confined} systems as well (Fig:\ref{fig:S_v_chi_fix-wall-diff-rho-combined}(B)
Similar to the 3D confined systems, the 2D \textit{x-confined} and \textit{xy-confined} systems also reveal that the particles accumulate near the wall and form the dense cluster close to the wall. Therefore we conclude that the accumulation of particles near the walls is a phenomena independent of the spatial dimensions.


\begin{figure}[]
    \centering
    \includegraphics[width=1.0\linewidth]{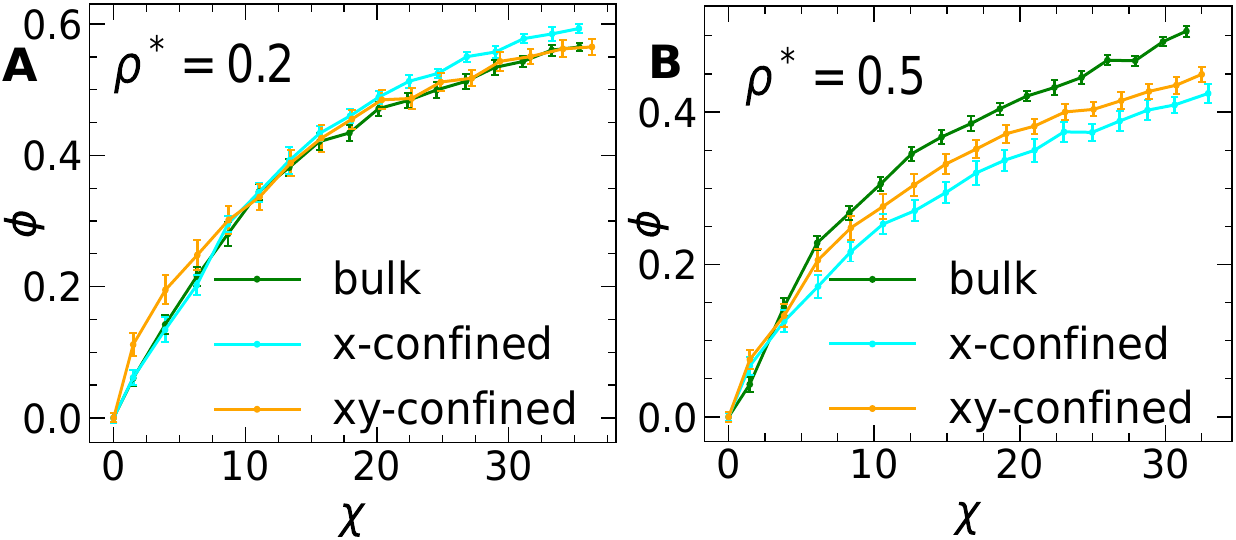}
    \caption{Effect of confinement on phase separation at a specific density in 2D - Plot of order parameter \(\phi\) vs activity \(\chi\) for three different confining cases: 2D bulk, \textit{x-confined} and \textit{xy-confined} case. A) For density $\rho^* = 0.2$, there is no observed effect of confinement. The order parameter $\phi$ almost takes the same value for bulk systems and confined systems (\textit{x-confined} and \textit{xy-confined}). B) For a density of $\rho^* = 0.5$, the effect of confinement is observed to reduce the phase separation. 
    }
    \label{fig:S_v_chi_fix-rho-diff-wall-combined}
\end{figure}


We also explore the effect of applying different confinements on the phase separation at fixed packing densities. We observe that at low densities such as $\rho^*=0.2$, (Fig:\ref{fig:S_v_chi_fix-rho-diff-wall-combined}(A)), the effect of confinement is not prominent in the phase separation, but at higher densities  $\rho^* = 0.5$, the phase separation again reduces with the insertion of wall in the system (Fig:\ref{fig:S_v_chi_fix-rho-diff-wall-combined}(B)), compared to 2D bulk systems.

As the spatial dimensions decrease from 3D to 2D, the reduced effective diffusivity of particles leads to the reduced thermal escape of the hot particles from the cold cluster. In addition, the interface between hot and cold clusters turns into a line in 2D systems from a surface in 3D systems. Also, the 2D systems can be viewed as a 3D confined system having an infinitesimal extent in one of the directions. These characteristics result in enhanced trapping of hot particles in cold clusters in two-dimensional systems. This phenomenon offers an explanation for the features observed in Fig:\ref{fig:S_v_chi_fix-wall-diff-rho-combined}(A) and (B). To quantify the trapping of hot particles, we define: 
$p_{cl}^{hot},p_{cl}^{cold} = (n_{cl}^{hot} / n_{cl}), (n_{cl}^{cold} / n_{cl})$, which denotes the fraction of hot and cold particles in the largest cluster with respect to the total particles in the cluster: \(n_{cl} = n_{cl}^{hot} + n_{cl}^{cold}\). Also \(f_{cl} 
 = n_{cl}/N\) denotes the fraction of particles taking part in the largest cluster and $f_{cl}^{hot} = (n_{cl}^{hot} / N_{hot})$ denotes the fraction of hot particles in the largest cluster, with respect to the corresponding hot system size.


 \begin{figure}[]
    \centering
    \includegraphics[width=1.0\linewidth]{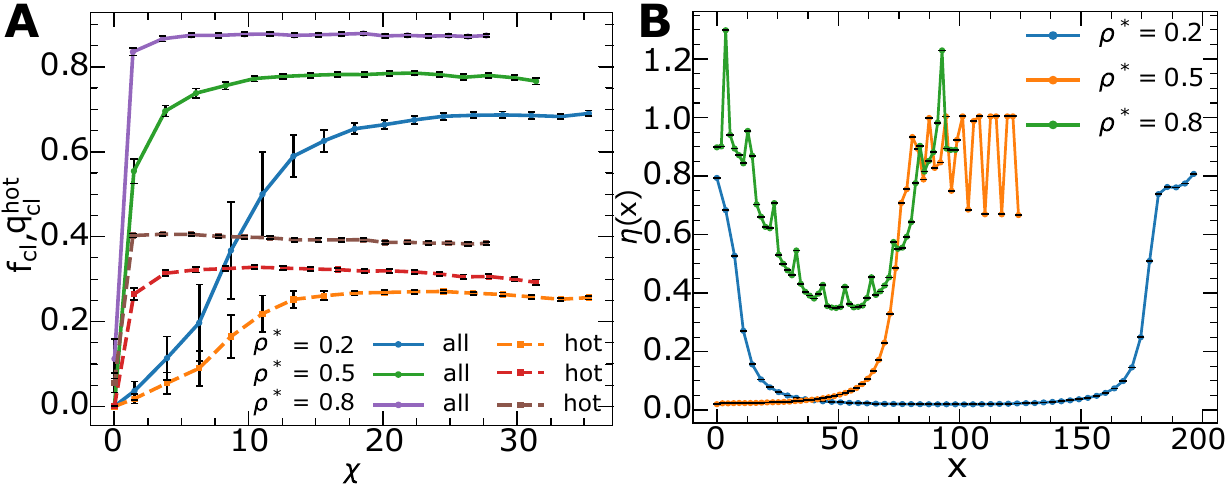}
    \caption{A) Composition of the largest cold cluster plotted against activity for the different densities considered: \(\rho^* = 0.2,0.5\) and \(0.8\) for the 2D bulk systems. The solid lines and dashed lines indicate the total number of particles \(f_{cl}\) and the trapped hot particles inside the cluster \( q_{cl}^{hot} (= f_{cl}^{hot}/2)\) respectively. The rise of the saturation values of $q_{cl}^{hot}$ with density \(\rho^*\) indicates the enhanced trapping of hot particles in cold cluster with density. For convenience in comparison, we took the value of \(1.1\sigma\) to be the cut-off distance for each density. B)The local packing fractions $\eta(x)$ for the non-equilibrium steady state for the \textit{x-confined} system, plotted against the distance along the wall $(x)$, for different initial equilibrium densities $\rho^* = 0.2,0.5$ and $0.8$. The phase-separated cold cluster gives rise to a high value of the local packing fraction, whereas the low values are due to the isotropic hot zone. The maximum value of the local density in cold cluster increases with the overall densities $\rho^*$.}
    \label{fig:density-dist-with-x-eta-all-pbc1}
\end{figure}


In fig. \ref{fig:density-dist-with-x-eta-all-pbc1}A) we show the behaviour of \(f_{cl}\) and \(q_{cl}^{hot} (= f_{cl}^{hot}/2)\) with \(\chi\) for 2D bulk systems. We observe that for a specific density \(\rho^*\), there is a rapid rise in the values of both the quantities with activity. This indicates that the size of the largest cluster as well as the number of trapped hot particles in the cluster rises with activity and then saturates to a certain value. We also note that the saturation value of \(q_{cl}^{hot}\) also grows with density, which is evidence of the enhanced trapping of hot particles as density is increased. This, in turn, explains why even for 2D bulk cases, we observe reduction in the phase separation with density (fig \ref{fig:S_v_chi_fix-wall-diff-rho-combined} A).

We also showed in Fig:\ref{fig:S_v_chi_fix-wall-diff-rho-combined}B that for confined systems there is a reduction in the phase separation with density. In addition to the enhanced trapping as discussed earlier, we also find that the variation of the local packing fractions in the simulation region also plays a role. We divide the \textit{x-confined} system into a number of slabs along x-axis and plot the local packing fraction $\eta(x)$ in each slab as a function $x$ (Fig:\ref{fig:density-dist-with-x-eta-all-pbc1}). The local packing fraction $\eta(x)$ is defined as $\eta(x) = \frac{N(x)\pi}{4L\Delta x}$, where $N(x)$  denotes the number of particles in a slab of width $\Delta x$ at a distance $x$. Due to the ordered structures in the cold domain near the wall, the local packing fraction attains a high value, whereas the dilute hot zone gives rise to the lower values of \(\eta(x)\). We notice that the maximum local packing fraction for the system increases with the initial density $\rho^*$. To give a quantitative idea, we note that for $\rho^* = 0.2$, the maximum local packing fraction $\eta_{max}(x) \approx 0.8$ (observed near $x \approx 0$), whereas for $\rho^* = 0.8$, $\eta_{max}(x) \approx 1.3$ (observed near $x \approx 10.0$). This means that as the total density $\rho^*$ increases, the virial pressure acting on the hot particles that are trapped inside the cold cluster also rises considerably. This prohibits their diffusion to the hot zone and keeps them trapped inside the cold cluster. This sustained trapping of hot particles in the cold cluster in turn reduces the phase separation between the active and passive particles for confined systems as the density increases.

\begin{table}[]
    \centering
\begin{tabular}{ |p {0.27\linewidth}|p{0.17\linewidth}|p{0.17\linewidth}|p{0.17\linewidth}| } 
 \hline
 \textbf{Periodicity} & \textbf{\% of total particles in largest cluster ($f_{cl}\times 100$)} & \textbf{\% of hot particles in cluster w.r.to cluster size ($p_{cl}^{hot}\times 100$)} & \textbf{\% of cold particles in cluster w.r.to cluster size ($p_{cl}^{cold}\times 100$)}  \\ \hline
 2D bulk & 76.59 \(\pm\) 0.78 &  38.18 \(\pm\) 1.27 & 61.82 \(\pm\) 0.85  \\ \hline 
 \textit{x-confined} & 89.97 \(\pm\) 0.33 & 46.54 \(\pm\) 0.43
 & 53.46 \(\pm\) 0.35\\  \hline
\textit{xy-confined} & 87.44 \(\pm\) 0.37 & 45.13 \(\pm\) 0.51
 & 54.87 \(\pm\) 0.37 \\ \hline
    \end{tabular}
    \caption{Table illustrating the enhanced trapping of hot particles in the cold cluster for confined systems in 2D at a specific density and activity $\rho^* = 0.5, T_h^* = 80.0$: The first column shows the different states of confinement. The second column presents the percentage of total particles (both hot and cold) $f_{cl}$ in the largest cluster. The 3rd column $p_{cl}^{hot}$ in the table shows that with confinement, more hot particles are trapped in the largest cluster.}
    \label{tab:table_confinement}
\end{table}

In general, for 2D bulk and confined systems, as the activity is increased, the size of the largest cluster ($f_{cl}$ or $n_{cl}$) increases. In Fig \ref{fig:S_v_chi_fix-rho-diff-wall-combined}B), we observe that for a specific density of \(\rho^* = 0.5\), there is a reduction in the extent of phase separation for confined systems compared to their bulk counterparts. This can be explained by analyzing \(p_{cl}^{hot}\) values for the bulk and confined systems at density \(\rho^* = 0.5\). For the confined system, at $\rho^* =0.5$, the fraction of hot particles trapped in the largest cluster also is greater compared to the bulk systems  ($(p_{cl}^{hot})_{x-confined} > (p_{cl}^{hot})_{bulk}$ at $\rho^* = 0.5, T_h^* = 80.0$, see table \ref{tab:table_confinement}). As a result, when $\rho^* = 0.5$, because the fraction of trapped hot particles is higher in confined cases than in periodic conditions, the order parameter is lower in confined systems than in bulk systems. In this way, the rise in hot particle trapping in cold clusters for confined systems explains the observed drop in phase separation order parameter for such confined systems in 2D at $\rho^* = 0.5$ (Fig:\ref{fig:S_v_chi_fix-rho-diff-wall-combined}B)

We conclude this section by briefly summarising the main results of this section. We observe that when the state of confinement is fixed and the density is varied, both the 2D bulk and confined systems show a reduction of phase separation between active and passive particles with density. This phenomenon is explained by analysing  the trapping of hot particles in the largest cluster in the system and also using the spatial variation of the local packing fractions of the confined systems 
(fig \ref{fig:density-dist-with-x-eta-all-pbc1}). Next, we observe that fixing the density and varying the state of confinement, at $\rho^* = 0.5$, the saturation value of $\phi$ is reduced for \textit{x-confined} and \textit{xy-confined} states compared to bulk systems ( Fig:\ref{fig:S_v_chi_fix-rho-diff-wall-combined}). We justify this observation via the analysis of enhanced trapping of hot particles in the cold cluster (table \ref{tab:table_confinement}). A general finding about 2D systems is the significant amount of trapping of hot particles in the cold cluster, which rises with density in both the bulk and confined scenarios. As a result, 2D binary mixtures at high densities can be thought of as the coexistence of a dense and dilute phase. This phenomenon is typical of 2D systems, and we address it in greater depth for the case of confinement on spherical surface in the following section.


\subsection{Effect of topology}

\begin{figure}[]
    \centering
    \includegraphics[width=1\linewidth]{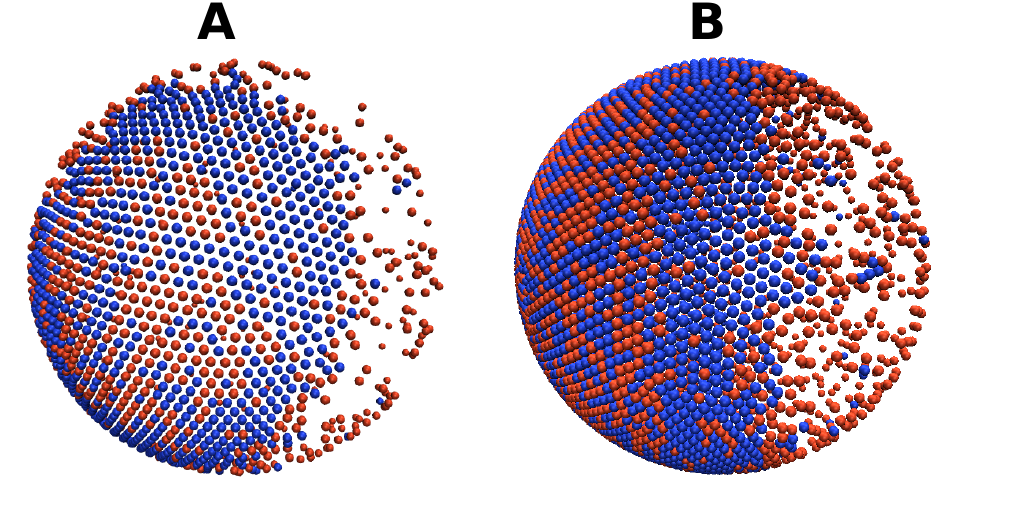}
    \caption{Snapshots of the non-equilibrium steady state of binary mixture confined on the surface of the sphere. The configuration of the system at $T_h^*=100.0, T_c^*=1.0, R=20.0\sigma$, A) $\eta=N/(16R^2) = 0.2$ and  B) $\eta = 0.6$. The configuration at $T_h^*=100.0$ shows a co-existence of dense and dilute regions on the spherical surface. The dilute region is hot-dominated, and the dense one is made up of both kinds of particles. The enhanced trapping with density can be visualized from the snapshots.}
    \label{fig:sphere-snapshots_eta0.6}
\end{figure}

\begin{figure}[]
    \centering
   \includegraphics[width=\linewidth]{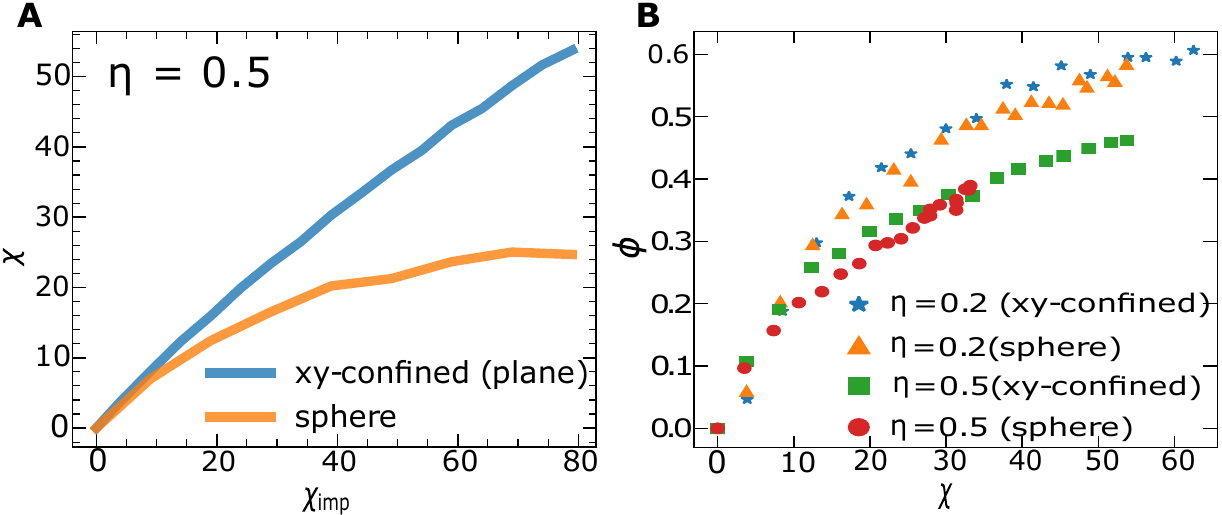}
   \caption{
   A) Variation of the observed activity \(\chi\) with the activity imposed by the thermostats \(\chi_{imp}\), for the \textit{xy-confined} system and particles confined on spherical surface case. For the case of 
 particles confined on spherical surface, due to the constraint equation imposed on the position and velocities of the particles, it is more difficult to numerically maintain the desired activity. In contrast, for the \textit{xy-confined} planar case, though the observed activity is less than the desired activity due to the collision between the particles, still this maintains the activity between the particles better than the spherical confinement. B) Comparison of order parameter \(\phi\) as a function of activity \(\chi\) between the \textit{xy-confined} and the case of particles confined on spherical surface. There is not much effect of the topological constraints on the phase separation phenomena, as is evident from the order parameter plots.
 }
   \label{fig:obs-temp-and-S-chiforplane-sphere}
\end{figure}

The \textit{xy-confined} state of confinement describes a system where the particles are confined from all directions. A spherical surface too corresponds to a similar finite-sized confined system. Such spherical systems will also reveal the effect of curvature on the various aspect of phase separation. Thus, to study the effect of topology and curvature, we simulate a binary system of LJ particles where particles are constrained to move on the surface of the sphere.  At every step, the positions and force vectors are projected back onto the local tangent plane to restrict the particles to the curved surface (eqn \eqref{sphere-pos-eq} and \eqref{sphere-vel-eq}). The equation of motion was integrated using the RATTLE algorithm \cite{paquay2016method} in LAMMPS. This algorithm yields results that, for suitably small integration timestep $\delta t$, are identical to those of actual motion on a spherical surface \cite{apaza2017brownian}. In line with the prevailing practice seen in many recent studies \cite{sknepnek2015active,janssen2017aging,henkes2018dynamical,ai2020binary,rajendra2023packing}, we have chosen to adhere to the convention of taking packing fraction as the measure of the concentration of the particles on the spherical surface. The packing fraction of particles on the spherical surface is defined as $\eta = N/(16R^2)$. Also, we restrain from studying the effect of the radius of curvature of the constraining sphere, and hence the simulations were carried out for a specific radius of the  sphere, $R = 20.0\sigma$. 
 Hence to change the packing fractions, the number of particles (\(N\)) is changed, keeping the radius $R$ fixed. 
 Also, the challenge of sustaining high activity for such spherical confinement, as elaborated further in subsequent discussions, prompts us to adopt an equilibrium temperature of \(T_c^* = T_h^* = 1.0\).
 The instantaneous equilibrium configuration for such a system at packing fraction $\eta = 0.6$, at $T_c^*=1.0$ is shown in fig \ref{fig:init}D. The particles are in a liquid state and are homogeneously dispersed over the surface of the sphere. Implementing the two-temperature model for the simulation of non-equilibrium behaviour, the observed steady-state configuration at $T_h^* = 80.0$ is shown in Fig:\ref{fig:sphere-snapshots_eta0.6}B ( the corresponding snapshot for \(\eta = 0.2\) is shown in Fig:\ref{fig:sphere-snapshots_eta0.6}A).   

The collision between the hot and cold particles helps in the exchange of energy between the particles, due to which we observe $\chi < \chi_{imp}$. In addition to that, as the mechanical constraints eq. \ref{sphere-pos-eq} and \ref{sphere-vel-eq} imposed on the particle position and velocity at each step of the simulation, it is harder to maintain the desired activity using the equilibrium thermostats for such spherical confinements, and therefore, we obtained a significantly lesser value of the observed activity $\chi$ compared to the desired activity $\chi_{imp}$ (fig \ref{fig:obs-temp-and-S-chiforplane-sphere}A). The definitions of $\chi,\chi_{imp}$ are given in section \ref{section:simulation details}.

At equilibrium ($T_h^* = T_c^* = 1.0, \eta = 0.6, R = 20.0\sigma$) the LJ particles form a liquid phase and they are homogeneously distributed over the surface of the sphere. When we increase the temperature of the hot particles to a high value ($T_h^* = 100.0, T_c^* = 1.0$), we observe the formation of a  dilute fluidlike phase, rich with hot particles, in co-existence with a solid-like dense region, which is made up of many hot particles trapped inside cold clusters (fig \ref{fig:sphere-snapshots_eta0.6} A, B). There is a circular band of cold particles (with finite width) forming at the interface between the two regions, which is necessary for the stability of the two co-existing phases. Since the particles are constrained on the surface of the sphere, defects are inevitable in the dense solid-like region. Therefore we calculated the defect structures for the solid-like phase and found that, apart from particles having co-ordination number 6, there are disclination defects where the number of nearest neighbours of a particle can be either 7 or 5 (see SI, fig 4). These two types of defect points often stay together forming dislocations, which are also "grain boundary scars", which separate the different domains within the crystalline structure. We note that earlier simulation works \cite{ai2020binary} of a binary mixture of self-propelled active and passive particles on spherical surface exhibit various other organisations and structures depending on the interplay between the polar alignment and rotational diffusion of the active particles. 

Next, We studied the effect of curved surfaces on the phase separation between the active and passive particles. In order to quantify the phase separation order parameter for such spherical surface confinement, we divide the whole surface of the  sphere into equal areas. For that, the azimuthal angle and the polar angle were divided accordingly. 
The details are discussed in section \ref{section:simulation details}. 

The results of order parameter calculations are shown in Fig:\ref{fig:obs-temp-and-S-chiforplane-sphere}B. We plotted the phase separation order parameter $\phi$ vs activity $\chi$ for two different packing fractions for spherical confinement, namely $\eta = 0.2$ and $\eta = 0.5$. To compare the results with a \textit{xy-confined} system, we also plotted the order parameter for the same at similar packing fractions(\(\eta = \frac{N\pi}{4L^2}, L = \)length of the plane) 
 (Fig:\ref{fig:obs-temp-and-S-chiforplane-sphere}B). We observe that: i) for the spherical surface confinement of the LJ binary mixture, the degree of phase separation between active and passive particles decreases as the packing fraction is increased. 
This is the same effect as observed in any other 2D system (bulk, \textit{x-confined} or \textit{xy-confined}) as discussed in detail in subsection \ref{subsec:Effect of confinement in 2 dimensional plane}. Therefore we conclude that the reduction in the extent of phase separation between active and passive particles as the packing fraction of the particles is increased- is a general phenomenon observed in 2D systems, independent of curvature or topology of the confining surface. 
ii) When we compare the order parameter values between the \textit{xy-confined} and spherically confined systems for a given packing fraction at different activities($\chi$), we find that the values are nearly identical. Hence, we conclude that the curvature has no substantial influence on the phase separation order parameter. 
Note that these comparisons are carried out by simulating the \textit{xy-confined} and spherically confined systems with equal surface areas of the confining geometry, to bring out solely the effect of constant curvature of the surface.

Although we showed that the order parameters are similar for the \textit{xy-confined} and spherically confined systems, there are certain aspects where these two systems differ. As discussed, one such aspect is the difference between the imposed and obtained activity ($\chi_{imp} - \chi$), which is larger for the spherically confined system (as shown in Fig:\ref{fig:obs-temp-and-S-chiforplane-sphere}(A)), the reason for which is discussed earlier.

A general observation for all considered 2D systems: bulk, planar confinement, and spherical confinement, is that as density increases, the phase separation reduces, which is the result of the increased trapping of hot particles in cold cluster. Hence, the systems can be also analysed in view of a dense-dilute phase co-existence, which is evocative of MIPS in 2D ABPs. The details are discussed in the next subsection.

\subsection{2-TIPS vs MIPS}


 \begin{figure}[]
    \centering
    \includegraphics[width=\linewidth]{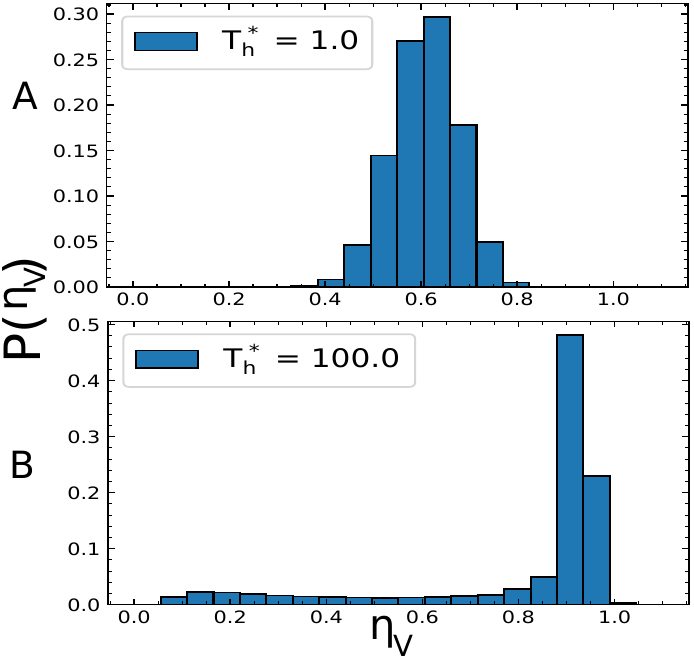}
    \caption{Distribution of the  Voronoi packing fraction \(\eta_V\) of the particles confined on the spherical surface at packing fraction $\eta = 0.6$. (A) In equilibrium, $T_h^* = 1.0, T_c^* = 1.0$, the histogram plot of the probability distribution of the Voronoi packing fraction associated with each of the particles shows a truncated Gaussian distribution, the centre of which is around $\eta_V = 0.6$. This denotes the equilibrium packing fraction $\eta = 0.6$. (B) As the temperature of the hot particles is increased to $T_h^* = 100$, we observe the distribution develops a very tiny peak around $\eta_V \approx 0.2$, and a larger peak around $\eta_V \approx 0.9$. This indicates that there is a co-existence between these two packing fractions at the steady state, indicating a MIPS-like behaviour. Although the curve does not show bimodality with equal strengths on the two modes.}
    \label{fig:areaDensity-dist}
\end{figure}

 \begin{figure}[]
    \centering
    \includegraphics[width=\linewidth]{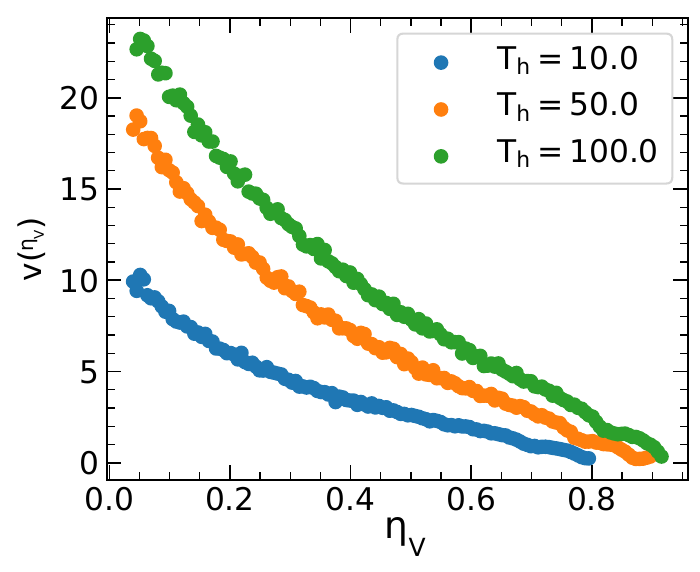}
    \caption{Average velocity of individual particles plotted against the average Voronoi packing fraction associated with each particle on the surface of the sphere. We observe that the velocity decreases with packing fraction (similar to MIPS), for each activity. Calculations are done at initial equilibrium packing fraction $\eta = 0.6, R = 20.0\sigma.$ }
    \label{fig:rho0.6_Th-all_area-velocity2}
\end{figure}


From the earlier sections we observe that, for 2D systems, the phase separation between the active and passive particles decreases at high densities. 
In fact, for the cases of spherical confinement, at high packing fractions ($\eta = 0.6$, fig \ref{fig:sphere-snapshots_eta0.6}B), we observe that the system phase separates between dilute and dense phases where a significant portion of the hot particles are trapped inside the domain of cold ones. 
For example, we found that $f_{cl} = 0.862 \pm 0.007$, $p_{cl}^{hot} = 0.434 \pm 0.010$ and $p_{cl}^{cold} = 0.566 \pm 0.006$, which means that $86.2\%$ of the total number of particles forms the largest cluster and $43.4\%$ of the  particles in the largest cluster  are hot particles (at $T_h^* = 100.0$).

Therefore, for the binary mixture of active and passive particles on the spherical surface, instead of further examining the phase separation between the two types of particles, an alternative approach is to consider the system as exhibiting the coexistence between a dense crystalline solid-like region and a dilute gaseous region. The dense region is composed of particles having small velocities and in the dilute region, the particle velocities are high. This effect is reminiscent of the Motility Induced Phase separation (MIPS) that has been observed for the purely active systems, which shows a co-existence between a dense and a dilute phase. We, therefore, have studied some of the aspects of MIPS in our cases, to have the comparison between 2-TIPS and MIPS phenomena.

In fig \ref{fig:areaDensity-dist}, we calculate the Voronoi packing fraction (which is defined as $\eta_V = 1/A_V$, where $A_V$ is the Voronoi area associated with each of the particles) for each of the particles at different activities. \nt{Note that, due to the intrinsic topology of the confinement, the local packing fractions are either undefined or carry significant statistical error due to a lesser number of data points. Hence, to quantify the packing fractions of the co-existing phases, we resort to the calculation of Voronoi packing fractions associated with each particle, and not the calculation of local packing fractions.} Fig:\ref{fig:areaDensity-dist}A) shows that at equilibrium (at $T_h^* = 1.0$ at $\eta = 0.6$), the distribution of $\eta_V$ is a Gaussian one (with finite extent), centered around the value $0.6$. At $T_h^* = 100.0$, the distribution of $\eta_V$ (Fig:\ref{fig:areaDensity-dist}B) shows a prominent peak around the value $0.9$, whereas another small peak is observed at around $\eta_V = 0.2$. This shows a co-existence of a dense and dilute phase on the spherical surface, at corresponding packing fractions $\eta_{1},\eta_{2} \approx 0.2,0.9$. Similar values of co-existent densities were obtained by Iyer $et. al$ \cite{iyer2022phase} in a system of ABPs on a spherical surface. Simulating a system of ABPs on a 2D plane, Redner $et. al.$ \cite{redner2013structure} showed, starting with an initial packing fraction $0.65$, the system demonstrates co-existence of two phases at packing fractions $\eta_1,\eta_2 \approx 0.2,1.2$. In addition to these observations, in Fig:\ref{fig:rho0.6_Th-all_area-velocity2}, we also show the average velocity of the particles and their dependence on their Voronoi packing fractions. We see a clear decreasing behaviour of the velocities with the Voronoi packing fractions, for different activities, which is a signature characteristic of MIPS.

Note that although the co-existence of a dense and dilute region of the particles on spherical surface, and the decreasing velocity of the particles with packing fraction are some of the similarities of the 2-TIPS on the spherical surface with MIPS, there are some distinct dissimilarities too. MIPS is observed in purely active systems, whereas we are observing it in a binary mixture of active and passive systems. Also, the phase co-existence is spontaneous for purely active systems and all the particles start with equal self-propelled velocity, but in 2-TIPS at high activity, there is an inherent velocity difference in the two types of particles, manifested by the two different temperatures.

\section{Conclusion:} \label{section:conclusion}
In this work, we examined the effect of confinement, geometry and topology on the phase separation phenomena in a system of a mixture of active and passive particles using the two-temperature model. We observe that:

i) For parallel wall confinement in 3D and for all 2D systems with or without confinement, the phase separation order parameter (although increasing with activity always) decreases as density is increased, which is completely opposite to the 3D bulk behaviour. 

ii) For a specific density, the parallel confinement can either enhance (3D, $\rho^* = 0.2$) or reduce (3D, $\rho^* = 0.8$;2D, $\rho^* = 0.5$), or almost unalter (2D, $\rho^* = 0.2$) the extent of phase separation between the active and passive particles compared to unconfined systems.
The interface between the phase-separated dense cold cluster and the dilute gaseous hot domain is always observed to form parallel to the confining wall. Enhancement of clustering due to confinement \cite{D2SM01012G} and accumulation of particles near the wall surface \cite{yang2014aggregation}, with experimental evidence \cite{Williams2022} has been reported earlier as well. 

iii) When the particles are confined inside of a spherical cavity, we observe the active and passive particles clearly phase separate and the type of phase separation changes with increasing the radius of the spherical cavity. When the radius of the cavity is small, the phase separation is observed in the radial direction, where the cold particles occupy the periphery of the confining sphere and the hot dilute zone is formed in the interior. As the radius of the spherical cavity increases, the bulk effects set in, and the phase separation is of Cartesian type. Spherical cavity confinement is always observed to enhance the phase separation compared to bulk cases at all observed densities. The confined steady states show a compression along the direction parallel to the wall in the cold zone, another feature completely opposite to the bulk (unconfined) cases.

iv) 2D systems in general shows enhanced trapping of hot particles in the cold cluster due to the reduced diffusivity of the particles. Thus, in the case of the binary mixture constrained on the surface of a sphere, our investigation encompassed not only the analysis of phase separation between active and passive particles but also the examination of a phase coexistence scenario involving a dilute gaseous phase and a densely packed crystalline domain. 

The dense crystalline phase on the surface of the sphere shows various defect structures which exist solely due to the topological constraint. These structures were compared with MIPS in 2D and various similarities as well as dissimilarities were figured out.

The future directions of this work may include the simulation of mixed systems on other topological surfaces and see whether they play any significant role in the phase separation phenomena or not. It will also be interesting to study the interplay between shape anisotropy and confinement by simulating mixtures of anisotropic particles under different geometric confinement. By changing the ratio between the active and passive particles, some critical activity can be determined for the different topological constraint cases, which would lead to some phase diagrams, which we can experimentally verify. These kinds of phase diagrams can be helpful in the experiments such as the behaviour of bacterial microswimmers on spherical liquid droplets and so on. Further analysis is required to study the dynamics of cold cluster formation both in the presence and absence of walls. The study of finite size effects and cluster dynamics along with stability analysis of the clusters will lead to a better understanding of the physics behind the phase separation processes in such binary mixtures in confined geometry and curved surfaces. We would like to publish further works along these lines in the upcoming days.

\section*{Author contributions}
Nayana Venkatareddy and Jaydeep Mandal: conceptualization, methodology, investigation, visualization, formal analysis, writing. Prabal K. Maiti: conceptualization, visualization, supervision, writing-review and editing. The manuscript was written through contribution from all authors.
\section*{Conflicts of interest}
There are no conflicts to declare.

\section*{Acknowledgements}
NV and JM thank MHRD, India for the fellowship. PKM thanks DST, India for financial support and SERB, India for funding and computational support. \nt{We also thank Profs. Chandan Dasgupta and Sriram Ramaswamy for the insightful discussions. } 


\balance


\bibliography{ref} 
\bibliographystyle{rsc} 

\end{document}


\maketitle
\begin{figure*}[h]
    \centering
    \includegraphics[width=1.0\linewidth]{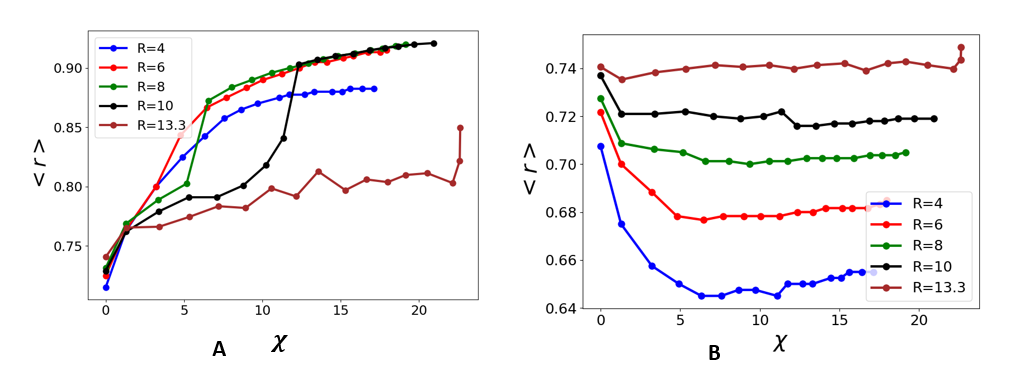}
    \caption{At $\rho^* = 0.8$ A) The plot of  mean radial position of the cold particles versus activity \(\chi\) for different values of $R$. The mean radial position \(<r>\) of cold particles increases with activity \(\chi\) for \(R=4\sigma\) to \(10\sigma\) indicating that higher number of cold particles move near wall at high activity leading to radial phase separation. For \(R=13.36\sigma\), \(<r>\) of cold particles does not reach a maximum value but saturates to a lower value implying that phase separation is not completely radial. B) The plot of  mean radial position of the hot particles versus activity \(\chi\) for different values of $R$. The mean radial position \(<r>\) of hot particles decreases with activity \(\chi\) implying that phase separation forces the hot particles to remain in interior of sphere away from wall.}
    \label{fig:mean_pos_hot-cold-allR}
\end{figure*}
\section{Mean radial position of hot and cold particles under spherical confinement}
 We have calculated the mean radial position of all cold particles \(<r>\) (normalized it by dividing it by radius R of the confining sphere) as a function of activity \(\chi\) for all radii \(R\) as shown in fig \ref{fig:mean_pos_hot-cold-allR}A.  If the value of \(<r>\) is closer to 1, it indicates that majority of cold particles are near the wall. For small radius of R= 4,6 and 8\(\sigma\), the mean radial position \(<r>\) of cold particles increases continuously with activity \(\chi\) and saturates to a value near the wall implying that the majority of cold particles are present near the periphery. For intermediate radius \(R=10\sigma\), the mean radial position \(<r>\) of cold particles reaches its maximum value only for high activity. For large radius \(R=13.36\sigma\), the mean radial position \(<r>\) of cold particles doesn't reach its maximum value even for high activities. So, smaller the radius of sphere and higher the activity, the higher the affinity of cold particles to be near the wall. Fig \ref{fig:mean_pos_hot-cold-allR}B shows plot of the mean radial position of all hot particles \(<r>\) as a function of activity \(\chi\) for all radius \(R\). For all radii we observe that the mean radial position \(<r>\) of hot particles (Fig \ref{fig:mean_pos_hot-cold-allR}B) decreases with activity and reaches a constant value as hot particles move into the interior away from the wall.\\
 \begin{figure*}
    \centering
    \includegraphics[width=1.0\linewidth]{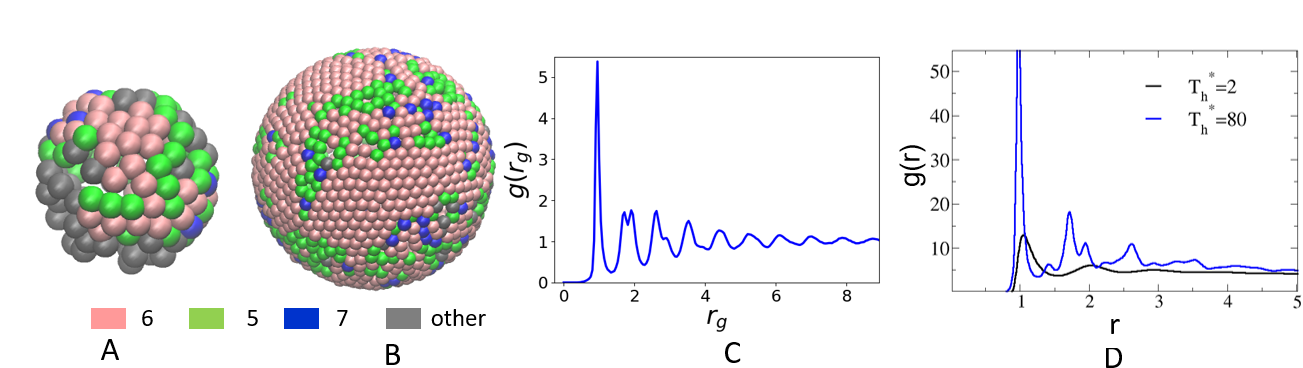}
    \caption{A) and B) Snapshots of cold clusters in the non-equilibrium system inside spheres of radius \(R=4\sigma\) and \(10\sigma\) respectively when \(\rho^*=0.8\) and \(T_h^*=80\). The particles are colored based on  the number of nearest neighbours. We can clearly see that fraction of particles with six neighbours increases as we move from \(R=4\sigma\) to \(10\sigma\). C) The radial distribution function of cold particles inside sphere of radius \(R=10\sigma\) when \(\rho^*=0.8\) and \(T_h^*=80\), calculated using geodesic distance \(r_g\) between particles shows crystal structure similar to 2D hexagonal lattice. D) The 3D radial distribution of cold particles inside sphere of radius \(R=13.36\sigma\) when \(\rho^*=0.8\) and \(T_h^*=80\) shows crystalline order similar to bulk phase separation.   }
    \label{fig:sphere_structure}
    \end{figure*}
    \begin{table}
    \centering
    \begin{tabular}{|c|c|c|c|}
    \hline
      R & 5 nn & 6 nn & 7 nn\\ \hline
 4 & 0.31 & 0.33 & 0.048\\ 
 6 & 0.33 & 0.48 & 0.076 \\
 8 & 0.23 & 0.64 & 0.088\\
 10& 0.183 & 0.71 & 0.087\\
 \hline
    \end{tabular}
    \caption{The table contains the fraction of particles in the cold cluster having 5,6 and 7 nearest neighbors (nn), for radius \(R= 4\sigma\) to \(10\sigma\) when \(\rho^*=0.8\) and \(T_h^*=80\).}
    \label{tab:1}
\end{table}
\section{Structure of phase-separated cold cluster under spherical confinement.}
 Mixture of hot and cold LJ particles in bulk at high activity phase separate into  crystalline cold clusters and a low-density region containing hot particles in gaseous phase. Cold particles of the phase separated system under spherical confinement also form two kinds of crystalline structures depending on the nature of phase separation: a 2D spherical shell of particles arranged in hexagonal lattice and a 3D cluster of cold particles with HCP and FCC arrangement. For radii in range of \(R=4\sigma\) to \(10\sigma\) at density \(\rho^*=0.8\) and \(T_h^*=80\), the 3D system of hot and cold mixture leads to the formation of 2D spherical shell of cold particles near the periphery of the sphere. We investigate the arrangement of cold particles in the shell by calculating the number of nearest neighbors (nn) for each particle. Table \ref{tab:1} gives the fraction of particles in the cold cluster having 5,6 and 7 nearest neighbors (nn), for radius \(R= 4\sigma\) to \(10\sigma\) when \(\rho^*=0.8\) and \(T_h^*=80\). The number of particles with 6 (5) nearest neighbors increases(decreases) as the radius of the confining sphere is increased. Fig \ref{fig:sphere_structure}A and B shows the snapshots of the cold cluster where the particles are colored based on the number of nearest neighbors when \(\rho^*=0.8\) and \(T_h^*=80\) inside spheres of radius \(R =4\sigma\) and \(10\sigma\) respectively. We can clearly see that \(R=10\sigma\) has larger domains of particles with six nearest neighbors compared to \(R=4\sigma\). The particles in cold cluster show  structure similar to that of an equilibrium system of particles on the surface of a sphere at high density showing hexagonal lattice with defects. So we calculate radial distribution function (RDF) g(\(r_g\)) of cold particles using geodesic distance \(r_g\) between the particles. The RDF g(\(r_g\)) at the geodesic distance \(r_g\) from the reference particles is calculated as  
\begin{equation}
    g(r_g)=\frac{N_{r_g}}{2 \pi R^2 sin\theta \Delta \theta \rho}
\end{equation}

where \(N_{r_g}\) is the number of particles in the spherical ring  at distance \(r_g=R\theta\) with area \(2 \pi R^2 sin\theta \Delta \theta\). Fig\ref{fig:sphere_structure}C shows RDF of cold particles inside sphere \(R=10\sigma\) when \(T_h^*=80\) using geodesic distance \(r_g\) between them. The g(\(r_g\)) shows splitting of second peak at \(\sqrt{3}\sigma\) and \(2\sigma\) resembling RDF in a 2D hexagonal lattice. \\
 Fig\ref{fig:sphere_structure}D shows 3D RDF g(r) of cold particles inside sphere \(R=13.36\sigma\) at  \(\rho^*=0.8\). The RDF of cold particles show liquid like order when \(T_h^*=2\) and crystalline order when \(T_h^*=80\) similar to bulk phase separation.

\section{ Structure of cold cluster of lj particles on spherical surface}
\begin{figure}[h]
    \centering
    \includegraphics[width=1.0\linewidth]{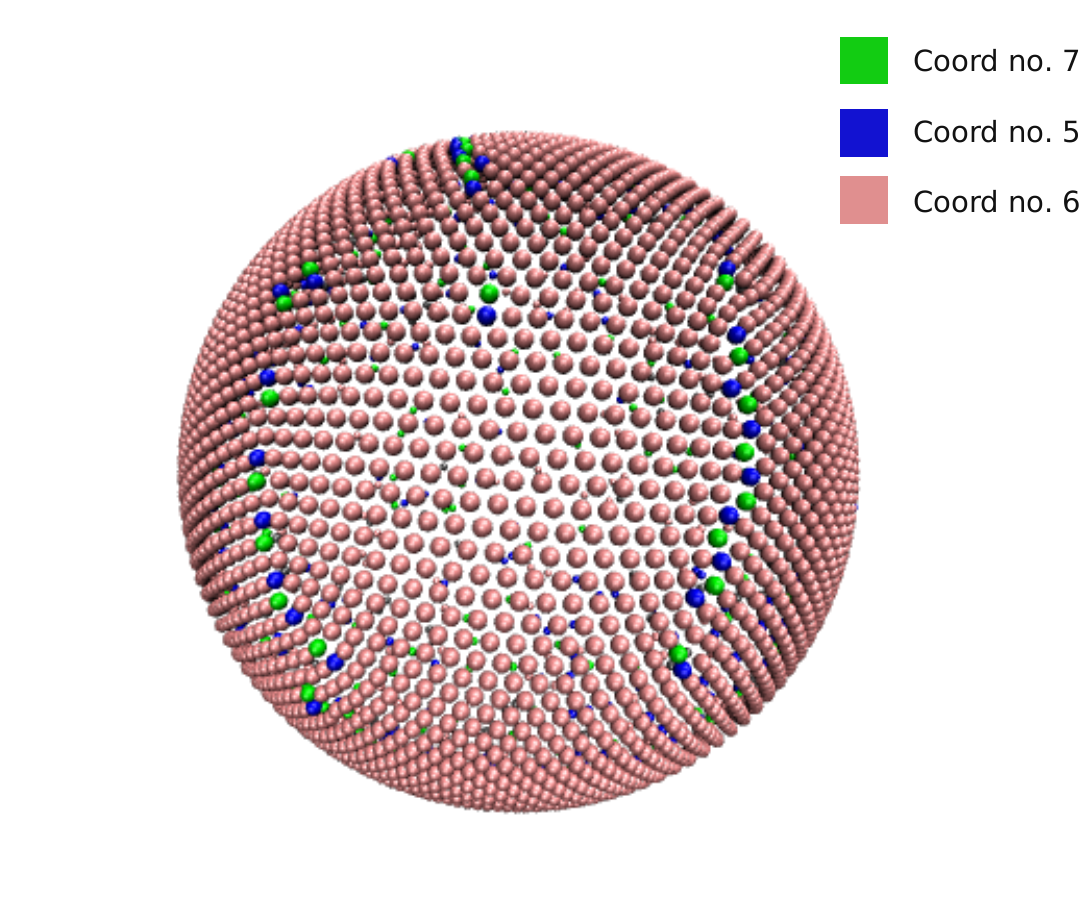}
    \caption{The defect points observed for the cold domain on the surface of the sphere. The points are color coded according to the number of their nearest neighbour, as shown in the figure. We observe that defects with co-ordination number 5 and 7 stay close to each other.}
    \label{fig:defects_with_colorcode}
\end{figure}

Perfect crystals cannot be formed on the surface of a sphere due to the curvature. The dense region of the non-equilibrium system on the spherical surface also shows the appearance of defects. The particles in the dense region arrange themselves in a hexagonal pattern. Therefore most particles have a coordination number of 6. But due to curvature, the co-ordination number of some point particles are other than 6 (namely 5 and 7). These two types of defects often stay together to form defect lines (fig \ref{fig:defects_with_colorcode})

\section{ Additional Figures}
\begin{figure}[h!]
    \centering
    \includegraphics[width=1.0\linewidth]{bulk pre.png}
    \caption{Variation of Normal pressure \(P_N\)=\(P_{xx}^*\) and tangential pressure \(P_T\)=\(P_{yy}^*\)=\(P_{zz}^*\) along \(\hat{x}\) direction perpendicular to interface between hot and cold particles when \(\rho^*=0.8\) and \(T_h^*=80\) in bulk simulation. We can see that in the cold zone, the tangential pressure is lower than normal pressure}
    \label{fig:bulkpr}
\end{figure}